\documentclass[12pt,reqno]{amsart}

\usepackage{amssymb}
\usepackage{graphicx}
\usepackage{accents}
\usepackage[cmtip,all]{xy}
\usepackage{verbatim}
\usepackage{tikz-cd}
\usepackage{stmaryrd}
\usepackage{wasysym}
\usepackage{graphicx}
\usepackage{mathrsfs}
\usepackage{enumerate}
\usepackage{geometry}
\usepackage{comment}

\DeclareMathAlphabet{\mathpzc}{OT1}{pzc}{m}{it}

\newtheorem{theorem}{Theorem}[section]
\newtheorem{lemma}[theorem]{Lemma}
\newtheorem{corollary}[theorem]{Corollary}

\theoremstyle{definition}

\theoremstyle{remark}

\numberwithin{equation}{section}

 \newcommand{\virgolette}{``}

\newcommand*{\defeq}{\mathrel{\vcenter{\baselineskip0.5ex \lineskiplimit0pt
                     \hbox{\scriptsize.}\hbox{\scriptsize.}}}%
                     =}

\newcommand\asim{\mathrel{%
  \ooalign{\raise0.1ex\hbox{$\sim$}\cr\hidewidth\raise-0.8ex\hbox{\scalebox{0.9}{$\scriptstyle{x}$}}\hidewidth\cr}}}

\newcommand{\mani}{\ensuremath{\mathpzc{M}}}
\newcommand{\manir}{\ensuremath{\mathpzc{M}_{red}}}
\newcommand{\stsheaf}{\ensuremath{\mathcal{O}_{\mathpzc{M}}}}

\newcommand{\beq}{\begin{equation}}
\newcommand{\eeq}{\end{equation}}
\newcommand{\bear}{\begin{eqnarray}}
\newcommand{\eear}{\end{eqnarray}}

\begin{document}
\begin{flushright}
\today
\par\end{flushright}
\vspace{1cm}

\title{On Forms, Cohomology, and BV Laplacians in Odd Symplectic Geometry}

\author{R.\ Catenacci}
\author{C.\ A.\ Cremonini}
\author{P.\ A.\ Grassi}
\author{S.\  Noja}

\email{roberto.catenacci@uniupo.it, carlo.alberto.cremonini@gmail.com}
\email{pietro.grassi@uniupo.it, sjmonoja@gmail.com}

\address{Dipartimento di Scienze e Innovazione Tecnologica - Universit\`a del Piemonte Orientale, Via T. Michel 11, 15121, Alessandria, Italy} 
\address{Dipartimento di Scienze e Alta Tecnologia - Universit\`a degli Studi dell'Insubria, via Valleggio 11, 22100 Como, Italy}
\address{Gruppo Nazionale di Fisica Matematica, InDAM, Piazzale Aldo Moro 5, 00185, Roma}
\address{Arnold Regge Center, Via P. Giuria 1, 10125, Torino, Italy}
\address{INFN Sezione di Torino, Via P. Giuria 1, 10125 Torino, Italy}
\address{INFN, Sezione di Milano, via G.Celoria 16, 20133 Milano, Italy}








\begin{abstract} We study the cohomology of the complexes of differential, integral and pseudo forms on odd symplectic manifolds taking the wedge product with the symplectic form as differential. We show that the cohomology classes are in correspondence with inequivalent Lagrangian submanifolds and that they all define semidensities on them. Further, we introduce new operators that move from one Lagragian submanifold to another and we investigate their relation with the so-called picture changing operators for the de Rham differential. Finally, we prove the isomorphism between the cohomology of the de Rham differential and the cohomology of BV Laplacian in the extended framework of differential, integral and pseudo forms.



\end{abstract}

\maketitle

\section{Introduction}

\noindent It is well-known and it has been discussed in several studies \cite{Manin, Voronov, Catenacci:2018xsv} that the geometry of forms and their complexes on supermanifolds proves to be remarkably richer compared to its counterpart on an ordinary purely commutative manifold. In order to have a coherent integration theory that accounts for the odd directions characterizing a supermanifold, differential superforms - the generalization of ordinary forms - need to be supplemented by the so-called integral forms. But this is not the end of the story: the space of forms on supermanifolds can further be enlarged by the so-called pseudo forms \cite{Witten}, which prove of crucial importance in many applications to string and quantum field theory, see for example \cite{Belo, Berkovits:2004px, NCPMR, CGNinf,CremoniniGrassi, CremoniniGrassi2, CremoniniGrassi3}. As we shall explain in the next section, the complexes of differential, integral as pseudo forms are distinguished and indexed by a number, called picture. \\
In this paper we focus on the geometry of forms on odd symplectic (super)manifolds, which provide the mathematical framework for a geometrization of the Batalin-Vilkovisky (henceforth BV) quantization, see the groundbreaking \cite{BV} and \cite{Schwarz}, the more recent \cite{Khuda1,Khuda2} or \cite{Mnev} for a thorough introduction to the subject. This very special class of supermanifolds are characterized by an odd 2-form $\omega$ which shares some properties with the ordinary symplectic form on a commutative even dimensional symplectic manifold and it can be used to define a Poisson-like bracket, called anti-bracket in the BV formalism. Our starting point, though, is the crucial observation of \v{S}evera \cite{Severa} that since $\omega$ is odd, it can be lifted to a nilpotent differential acting on forms (by wedge product), and it therefore makes sense to study its cohomology. In particular, we study the cohomology of $\omega$ for the complexes of differential, integral and pseudo forms on a generic odd symplectic supermanifolds and we show that classes in this cohomology are related to inequivalent Lagrangian submanifolds in the odd symplectic supermanifolds, see again \cite{Schwarz}.
The contact with BV formalism is made showing that all of these classes define semidensities on the odd symplectic supermanifold, where the BV Laplacian acts \cite{Khuda1}. Further, we introduce new operators which relate different cohomology classes, which is equivalent to jump from one Lagrangian submanifold to another. This kind of operators are known in string theory as picture changing operators (PCO's); they are usually introduced in connection with the de Rham differential \cite{Belo, Catenacci:2016qzd, Catenacci:2018xsv}: we show that these newly constructed PCO's, acting on the cohomology of $\omega$, are related to the PCO's of the de Rham differential by similarity transformations. \\
Another simple, yet amazing, observation in \cite{Severa}, is that the de Rham differential $d$ and the odd symplectic form $\omega$ (anti)commute, making the de Rham complex into a double complex having $d$ and $\omega$ as differentials. This remarkable fact implies that the BV Laplacian \cite{Khuda} arises naturally - and invariantly - as the last non-zero differential in the related spectral sequence starting with $\omega.$ We broaden this analysis to the extended de Rham complex comprising differential, pseudo and integral forms. Further we study directly the cohomology of the BV Laplacian, by showing its homotopy - whose explicit form does not appear in the literature to the best knowledge of the authors. Finally we compare the two spectral sequences - the one starting with $\omega$, which computes the cohomology of the BV Laplacian and the one starting with $d$ which computes the de Rham cohomology. We realize explicitly the isomorphism between these two cohomology in two ways, one of which makes use of the previously introduced picture changing operators for $\omega$ and $d$.\\
Finally, the last part of the paper is more speculative. Using some of the ingredients introduced in the paper, we argue an analogy between ordinary complex geometry and odd symplectic geometry: within this framework we show how to write down an action for a Kodaira-Spencer-type theory \cite{BCOV} in the context of odd symplectic supermanifolds. This result is evocative, but it still requires a deep investigation, which we leave to future works.

\section{Basics of Supermanifolds and Odd Symplectic Geometry}

\noindent Before we start, we fix our notations. We let $M$ and $\mani$ be an ordinary smooth manifold and a smooth supermanifold respectively. If $M$ is of dimension $n$, locally, we describe it by means of a chart $(U, x^i)$ for $i= 1, \ldots, n$ and $U$ and open set in the topological space underlying $M$. Likewise, if $\mani$ is of dimension $n|p$, locally, we describe it by means of a chart of both even $x$ and odd coordinates $\theta$ given by $(U, x^i | \theta_\alpha )$ for $i=1, \ldots, n$ and $\theta_\alpha = 1, \ldots, p.$ As it is customary, we will call $\manir$ the \emph{reduced space} - \emph{i.e.}\ the ordinary manifold - underlying $\mani.$  Also, given a supermanifold $\mani$ of dimension $n|p$ we denote $\mathcal{T}_\mani$ and $\Pi \mathcal{T}_\mani$ the tangent bundle and its parity changed version: notice that these have rank $n|p$ and $p|n$ respectively, being they locally generated by the derivations $\{ \partial_{x^i} | \partial_{\theta_\alpha} \}$ and $\{\pi \partial_{\theta_\alpha} | \pi \partial_{x^i}\}$ respectively. The dual of these bundles are given by $\Omega^1_{\mani, ev} \defeq \mathcal{T}_\mani^\ast$ and $\Omega^1_{\mani, odd} = \Pi \mathcal{T}_\mani^\ast$: here we will only use the second one, which is of rank $p|n$ and locally generated by $\{ d\theta_\alpha | d x^i \}$ and we will denote it simply as $\Omega^1_{\mani}$ or $\Pi \mathcal{T}^\ast_\mani.$ The Berezinian bundle of a supermanifold is defined to be $\mathcal{B}er (\mani) \defeq \mathcal{B}er (\Pi \mathcal{T}^\ast_\mani)^\ast:$ this is of parity $(p+q)\mbox{mod}\, 2.$ The reader is invited to refer to \cite{Fioresi, Manin} or the recent \cite{CNproj, CNR} for details.

\noindent Also, in order to relate the theory of forms with integration theory over supermanifolds, the ordinary de Rham complex needs to be extended by the so-called \emph{integral} and \emph{pseudo-forms}, which supplement ordinary differential forms in $\Omega^\bullet_\mani = S^\bullet \Omega^1_\mani$. Given a supermanifold $\mani$ of dimension $n|q$, these \virgolette generalized'' forms are labelled by a number, called \emph{picture} $p$, for $p=1, \ldots, q$, and for this reason we will denote this extended de Rham complex by $\Omega^{\bullet |p}_\mani.$ Usual differential superforms are sections of $\Omega^{\bullet|0}_\mani$ and control integration over submanifolds of codimension $k|q$ in $\mani$, integral forms have maximal picture, \emph{i.e} are sections of $\Omega^{\bullet|q}_\mani$ and control integration over sub-supermanifolds of codimension $k|0$, whilst pseudo forms are section of $\Omega^{\bullet |p}_\mani$ for $p \neq \{ 0, q\}$ and control integration over more general sub-supermanifolds of codimension $k|l$ for $l \neq \{ 0, q\}.$ In this framework forms in $\Omega^{\bullet | p}_\mani$, for $p=0, \ldots, q$ can be seen as generalized functions over $\mbox{Tot} (\Pi \mathcal{T}_\mani)$ - this point of view is highlighted in \cite{Witten}. Differential forms, in picture $p=0$, are characterized by a \emph{polynomial} dependence of all of the even fiber coordinates $d\theta$'s, pseudo forms with picture $0 < p  < q $ are characterized by Dirac-delta \emph{distributional} dependence on $p$ out of the $q$ $d\theta$'s, while the remaining $q-p$ are allowed a polynomial dependence. Finally, integral forms, having maximal picture $p = q$, have distributional dependence on all of the $d\theta$'s. Notice also that any $\Omega_\mani^{r|p}$ is a $\mathcal{O}_\mani$-module \emph{and} also $\Omega^{\bullet|0}_\mani$-modules, \emph{i.e.} any pseudo and integral forms can be multiplied not only by functions but also by differential forms in $\Omega^{n|0}_\mani$ for any $n$. The de Rham differential can be generalized as to operate in this extended framework, so that $d : \Omega^{k|p}_\mani \rightarrow \Omega^{k+1|p}_\mani,$ thus giving a \virgolette stack'' of $q$ complexes $\Omega^{\bullet| p}_\mani$ for $p=0, \ldots q$. This is represented by the following diagram
\bear \label{12}
\xymatrix{
  & 0 \ar[r] & \Omega^{0|0}_\mani \ar@/^.6pc/[d]^{\mathbb{Y}} \ar[r] & \Omega^{1|0}_\mani \ar@/^.6pc/[d]^{\mathbb{Y}} \ar[r] & \ldots \ar@/^.6pc/[d]^{\mathbb{Y}}\ar[r] & \Omega^{n|0}_\mani \ar@/^.6pc/[d]^{\mathbb{Y}} \ar[r] & \Omega^{n+1|0}_\mani \ar@/^.6pc/[d]^{\mathbb{Y}} \ar[r] & \ldots \\
  \ldots \ar[r] & \ar@/^.6pc/[d]^{\mathbb{Y}}  \Omega^{-1|1}_\mani \ar[u]^{\mathbb{Z}} \ar[r] & \Omega^{0|1}_\mani \ar@/^.6pc/[d]^{\mathbb{Y}} \ar@/^.6pc/[u]^{\mathbb{Z}} \ar[r] & \Omega^{1|1}_\mani \ar@/^.6pc/[d]^{\mathbb{Y}} \ar[r] \ar@/^.6pc/[u]^{\mathbb{Z}}& \ldots \ar@/^.6pc/[d]^{\mathbb{Y}} \ar@/^.6pc/[u]^{\mathbb{Z}}\ar[r] & \ar@/^.6pc/[d]^{\mathbb{Y}}\Omega^{n|1}_\mani \ar@/^.6pc/[u]^{\mathbb{Z}}\ar[r]\ar[r] & \ar@/^.6pc/[u]^{\mathbb{Z}}\ar[r] \Omega^{n+1|1}_\mani \ar@/^.6pc/[d]^{\mathbb{Y}} \ar[r] & \ldots \\
\ldots \ar[r] & \ldots \ar@/^.6pc/[u]^{\mathbb{Z}}\ar[r] \ar@/^.6pc/[d]^{\mathbb{Y}} & \ar@/^.6pc/[d]^{\mathbb{Y}}  \ldots  \ar@/^.6pc/[u]^{\mathbb{Z}} \ar[r] & \ldots \ar@/^.6pc/[d]^{\mathbb{Y}} \ar@/^.6pc/[u]^{\mathbb{Z}} \ar[r] & \ldots \ar@/^.6pc/[d]^{\mathbb{Y}} \ar[r] \ar@/^.6pc/[u]^{\mathbb{Z}}& \ldots \ar@/^.6pc/[d]^{\mathbb{Y}} \ar@/^.6pc/[u]^{\mathbb{Z}}\ar[r] & \ldots \ar[d]^{\mathbb{Y}} \ar@/^.6pc/[u]^{\mathbb{Z}}\ar[r] & \ldots  \\
 \ldots  \ar[r] & \ar@/^.6pc/[u]^{\mathbb{Z}} \Omega^{-1|q}_{\mani} \ar[r] & \ar@/^.6pc/[u]^{\mathbb{Z}} \Omega^{0|q}_{\mani} \ar[r] & \ar@/^.6pc/[u]^{\mathbb{Z}} \Omega^{1|q} \ar[r] & \ldots \ar@/^.6pc/[u]^{\mathbb{Z}} \ar[r] & \ar@/^.6pc/[u]^{\mathbb{Z}}  \Omega^{n|q}_\mani \ar[r] & 0
}
\eear
The (local) operators $\mathbb{Y}$ and $\mathbb{Z}$ moving vertically between complexes labelled by different pictures are called \emph{Picture Changing Operators}, and have made their first appearance in a geometric context in \cite{Belo}. Further examples of this kind of operators will be provided in the following.\\




\subsection{Odd Symplectic Supermanifolds} In this paper, in particular, we will deal with a special class of supermanifolds, the so called \emph{odd symplectic supermanifolds.} These are defined as pairs $(\mani, \omega)$, where $\mani$ is a supermanifold and $\omega \in \Omega^{(2|0)}_\mani$ is $d$-\emph{closed}, \emph{i.e.}\ $d\omega = 0$, \emph{odd} and \emph{non-degenerate}, \emph{i.e.}\ it can be represented as $\omega \lfloor_{U} = \sum_{i, \alpha=1}^n \omega_{i\alpha} (x, \theta) dx^i \wedge d\theta_\alpha $ in local coordinates $(U, x^i| \theta_\alpha) $, with $\omega_{i\alpha} $ an \emph{invertible} matrix taking values on $\mathcal{O}_\mani  (U).$ Notice that requiring $\omega$ to be non-degenerate constraints $\mani$ to have the same number of even and odd dimensions, \emph{i.e.} an odd symplectic supermanifolds is always of dimension $n|n$ for some $n.$ These class of supermanifolds has been fully characterized by Schwarz in \cite{Schwarz}.
\begin{theorem}[Schwarz \cite{Schwarz}] \label{11} Let $ ( \mani, \omega )$ be an odd symplectic supermanifold, with reduced manifold $\manir.$ Then one has
\begin{enumerate}
\item in the neighborhood $U_p$ of any point $p$ in $\manir$ there exist a system of \emph{local} coordinates $(U_p, x^i | \theta_i)$ such that $\omega = \sum_{i=1}^n dx^i \wedge d \theta_i$; 
\item there exists a \emph{global} symplectomorphism $\varphi : (\mani, \omega) \rightarrow (\mbox{\emph{Tot}} ( \Pi \mathcal{T}^\ast_{\manir}), \omega_{std}),$ where $\omega_{std} = \sum_i dx^i \wedge d\theta_i$. 
\end{enumerate}
\end{theorem} 
\noindent The first point in the above theorem proves the existence of \emph{Darboux coordinates} also in this supersetting, whilst the second point provides a global description of the geometry of odd symplectic supermanifolds: up to symplectomorphisms, \emph{odd symplectic supermanifolds are all total spaces of odd cotangent bundle of some ordinary manifold $M$}. It is therefore not restrictive to limit to this kind of geometries only.\\
More in details, an odd symplectic supermanifold $\mbox{Tot} (\Pi \mathcal{T}^\ast_M) \defeq \Pi \mathcal{T}^\ast_M \stackrel{\pi}{\rightarrow} M$ will be characterized as follows: its underlying reduced manifold coincides with $M$ and its structure sheaf is given by $(S^\bullet \Pi \mathcal{T}_M)^\ast$, \emph{i.e.}\ functions of $\mbox{Tot} (\Pi \mathcal{T}^\ast_M)$ are polynomial functions over the fibers $\mathcal{T}_{M, p}^\ast$. This means that if $(U, x^i)$ is a chart over $M$, then $(U, x^i | \theta_i)$ is a chart over $\mbox{Tot} (\Pi \mathcal{T}^\ast_M)$, where the odd coordinates $\theta_i$ are given by $\theta_i \defeq \partial_{dx^i}.$ Notice that this implies that in an intersection $U \cap V$, of charts $(U, x^i|\theta_i)$ and $(V, y^i|\psi_i)$ one has that $\theta_i = \sum_{j=1}^n (\partial_{x^i} y^j )\psi_j$, \emph{i.e.}\ the odd coordinates $\theta_i$ transform with the (transpose) inverse Jacobian of the change of coordinates $x^i \mapsto y^i $ over the base $M$.\\
Let us now look at forms over the odd symplectic supermanifold, that for notational convenience we will write as $ \Pi \mathcal{T}^\ast M \defeq \mbox{Tot} (\Pi \mathcal{T}^\ast_M)$. On a geometrical ground, $\Omega^1_{\Pi T^\ast M}$ can be seen to be an extension of $\displaystyle \Omega^1_{\Pi T^\ast M / M}$ by $\pi^\ast \Omega^1_M$, where $\pi : \Pi \mathcal{T}^\ast M \rightarrow M$ via the (canonical) short exact sequence 
\bear \label{canses}
\xymatrix{
0 \ar[r] & \pi^\ast \Omega^1_M \ar[r] & \Omega^1_{\Pi \mathcal{T}^\ast M} \ar[r] & \Omega^1_{\Pi \mathcal{T}^\ast M / M} \ar[r] & 0.
}
\eear
Since we are working over a smooth manifold $M$, these vector bundles are fine sheaves (of $\mathcal{O}_M$-modules) and therefore acyclic. It follows that, in particular, $Ext^1 ( \Omega^1_{\Pi T^\ast M / M}, \pi^\ast \Omega^1_M) \cong 0$, and the short exact sequence \eqref{canses} is (non-canonically) split, \emph{i.e.} $\Omega^1_{\Pi T^\ast M} \cong \pi^\ast ( \Omega^1_M  \oplus \Omega^1_{\Pi T^\ast M / M} )$ (notice, incidentally, that this is equivalent to say that there exists a non-canonical reduction of the structure group $Sp(n|n) $ preserving $\omega$ to $GL(n) \times GL (n)$). From the expressions of the transition functions for the local coordinates $x^i | \theta_i$ of the odd symplectic manifold one can then easily find the transition functions for the local generator $d\theta_i$ and $dx^i$ of $\Omega^1_{\Pi \mathcal{T}^\ast M}$: the splitness of the \eqref{canses} implies that $dx^i = \sum_{j=1}^n dy^j (\partial_{y^j} x^i)$ and that $d\theta_i = \sum_{j=1}^n (\partial_{x^i} y^j) d \psi_j$, which in turn identifies $\Omega^1_{\Pi \mathcal{T}^\ast M / M} \cong \pi^\ast \mathcal{T}_M$ \footnote{Indeed, notice that from the transformation property of $\theta_i$ one would expect $d \theta_i = \sum_{j=1}^n (\partial_{x^i} y^j ) d\psi_j + \sum_{j=1}^n d (\partial_{x^i} y^j )\psi_j $, in agreement with $\Omega_{\Pi \mathcal{T}^\ast\mani}$ being an extension of vector bundles. Since \eqref{canses} splits, charts can be chosen as to systematically drop the second term, hence $d \theta_i$ transforms as $\theta_i$. This problem could be tackled as well by considering a covariant derivative instead of the exterior derivative $d$.}. This decomposition makes very easy to prove an important property of odd symplectic supermanifolds, namely if $\Pi \mathcal{T}^\ast M$ is an odd symplectic supermanifold then $\mathcal{B}er ( \Pi \mathcal{T}^\ast M) \cong \pi^\ast (\det (M))^{\otimes 2}$, where $\pi^\ast$ is the pull-back of $\pi: \Pi \mathcal{T}^\ast M \rightarrow M$, the structure map of the fibration and $\det (M)$ is the canonical bundle of $M$.

\noindent Notice that this can also be seen from the interpretation of the Berezininan bundle via (dual of the) Koszul complex related to the tangent bundle of a supermanifold. Without giving further details, given a generic supermanifold $\mani$, which is not necessarily an odd symplectic supermanifold, the only non-trivial representative in the related cohomology is given by $[dx^1 \ldots dx^n \otimes \partial_{\theta_1} \ldots \partial_{\theta_n}] \in \mathcal{E}xt^{n}_{S^\bullet \mathcal{T}_\mani} (\mathcal{O}_\mani, S^\bullet \mathcal{T}_\mani)$: this class transforms with the (inverse) Berezinian of the Jacobian of the change of coordinates, \emph{i.e.}\ as a generating section of $\mathcal{B}er(\mani)$. It follows that one gets a cohomological characterization of the Berezinian via the Koszul complex construction, which non-trivially generalizes the analogous construction for $\det (M)$ from the ordinary Koszul complex (this is hinted in \cite{Manin}). It is then easy to verify, using the transition functions given above (and their dual) that in the case of an odd symplectic supermanifold the class $[dx^1 \ldots dx^n \otimes \partial_{\theta_1} \ldots \partial_{\theta_n}]$ transforms with the second power of the canonical bundle over $M,$ indeed in the (class of the) tensor product above the bit $\partial_{\theta_1} \ldots \partial_{\theta_n}$ transform exactly as $dx^1 \ldots dx^n.$

\section{The Cohomology of the Odd Symplectic Form}

\noindent As can be easily seen from the previous considerations, the odd symplectic form $\omega$ is invariant under generic change of coordinates and, being odd, it is \emph{nilpotent}. It follows that multiplication by $\omega$ can be seen as an odd nilpotent morphism from $\Omega^{\bullet | p}_\mani$ to itself since all of these are sheaves of $\Omega^{\bullet}_\mani = \Omega^{\bullet | 0}_\mani$-modules. In other words, given the extended de Rham complex as above, we have another complex $(\Omega^{\bullet | p}_\mani, \omega)$, which is characterized by the multiplication by $\omega$, \emph{i.e.} $\alpha \mapsto \omega \alpha $, as nilpotent differential. In \cite{Severa}, \v{S}evera showed that $(\Omega^{\bullet | 0}_\mani, d, \omega)$ forms a bicomplex and proved that the cohomology $H_\omega (\Omega^{\bullet |0}_\mani)$ is (naturally) isomorphic to \emph{semidensities} on $\mani,$ see \cite{Severa} \cite{Khuda}. \\

\noindent In this section we start by revising the result in \cite{Severa} and we extend it to the whole extended de Rham complex $\Omega^{\bullet |p}_\mani$. Let us look at an ordinary differential superform at picture $p=0$ of any degree, we call it $\alpha_{0}$. Notice that $\alpha_0$ is locally of the form $\alpha_0 = f(x, \theta) (dx^1)^{\epsilon_1} \ldots (dx^n)^{\epsilon_n} d \theta_1^{k_1} \ldots d \theta_n^{k_n}$ for $k_i \in \mathbb{N}_0$ and $\epsilon_i \in \{ 0,1\}.$  In this case a homotopy for $\omega$ is given by the \virgolette counting'' operator $\mathpzc{h} : \Omega^{\bullet | 0}_\mani \rightarrow \Omega^{\bullet - 2 | 0}_\mani$, locally given by $\mathpzc{h} \defeq \sum_{i=1}^n \iota_i \iota^i = \sum_{i=1}^n \partial_{d\theta_i} \partial_{dx^i}$. An easy calculation yields
\begin{align}
(\mathpzc{h} \omega + \omega \mathpzc{h}) \alpha_{0} = \left ( n + d\theta_i \iota^i - dx^i \iota_i \right ) \alpha_{0} = \left ( n + \deg_{d\theta} (\alpha_{0}) - \deg_{dx} (\alpha_{0}) \right )\alpha_{0},
\end{align}
where $\deg_{d\theta} (\alpha_0)$ is the even degree, \emph{i.e.} $\deg_{d\theta} = \sum_{i=1}^n k_i$ of $\alpha_0$ and $\deg_{dx} (\alpha_0)$ is the odd degree, \emph{i.e.}\ $\sum_{i = 1}^n \epsilon_i$ of $\alpha_0.$ It follows that the only instance in which the homotopy fails is $\deg_{d\theta} (\alpha_0) = 0$ and $\deg_{dx} (\alpha_0) = n$, that is for elements in $\Omega^{n|0}_\mani$ of the form $\alpha_0 = f (x,\theta) dx^1 \ldots dx^n$. Notice that this elements is indeed a cycle and therefore defines a class in cohomology. Using the above transformation properties, it is straightforward to verify that elements in this cohomology class transform as a semidensities \emph{i.e.} like sections in $\mathcal{B}er (\mani)^{\otimes 1/2}$: in the case of an odd symplectic supermanifold $\mani = \Pi \mathcal{T}^\ast M$ one has that $\det (M) \cong\mathcal{B}er (\mani)^{\otimes 1/2}$, which proves the claim.\\

\noindent Let us now move to maximal picture $p=n$ and consider the case of integral forms. Before we go, we recall that a generic integral form $\alpha_q$ can be written locally in the form $\alpha_q = f(x,\theta) (dx^1)^{\epsilon_1} \ldots (dx^n)^{\epsilon_n} \delta^{(\ell_1)} (d\theta_1) \ldots \delta^{(\ell_n)} (d\theta_n)$, again for $\epsilon_i \in \{ 0,1\}$ and $\ell_i \in \mathbb{N}_0$ and since there are all of the $\delta (d\theta)$'s there cannot be any $d\theta$'s. Once again, the homotopy $\mathpzc{h}$ introduced above does the job, and one gets
\bear
(\mathpzc{h} \omega + \omega \mathpzc{h}) \alpha_n =  \left ( n - \deg_{dx} (\alpha_n) - \deg_{\delta (d\theta)} (\alpha_n) - n \right ) \alpha_n,
\eear
where we have further defined $\deg_{\delta (d\theta)} (\alpha_n) = \sum_{i=1}^n \ell_i$. It follows that the homotopy fails in the case that $\deg_{dx} (\alpha_n)$ and $\deg_{\delta (d\theta)} (\alpha_n) $ are both zero, corresponding to the integral form of degree zero given by $\alpha_n = f (x, \theta) \delta (d\theta_1) \ldots \delta (d\theta_n) \in \Omega^{0|n}_{\mani},$ which is readily verified to be a cycle. It is once again not hard to see from this expression - using the transformation properties of the delta's and of the $d\theta$'s - that it transforms as a section of $\det (M)$, \emph{i.e.}\ it is once again a semidensity. Another way to see this, is to exploit the different representation of integral forms as sections of $\mathcal{B}er (\mani) \otimes S^\bullet \Pi \mathcal{T}_\mani$, see \cite{Manin} and \cite{Catenacci:2018xsv} for explicit example of this correspondence. Up to the function $f (x, \theta)$, the form $\alpha_n$ corresponds to the section $ \mathcal{D} (x,\theta)  \otimes \pi \partial_{x^1} \ldots \pi \partial_{x^n}$, where $\mathcal{D} (x,\theta)$ is a generating section of $\mathcal{B}er (\mani) \cong \det (M)^{\otimes 2}$ and $\pi \partial_{x^1} \ldots \pi \partial_{x^n}$ is a section in $S^n \Pi \mathcal{T}_\mani$, which is totally antisymmetric and it transforms as a section of the anticanonical bundle $\det (M)^{\otimes -1}$. Just by looking at the above transformation properties, one has that that $\alpha_n$ transforms as a section in $\det ({M})^{\otimes 2} \otimes \det (M)^{\otimes -1} \cong \det (M) \cong \mathcal{B}er (\mani)^{\otimes 1/2}.$\\ 


\noindent Finally, let us deal with the case of pseudo forms, which - as usual - are the most delicate. We recall that a generic pseudo form $\alpha_p$ at picture $0 < p < n$ is locally written as $\alpha_p = f (x, \theta) (dx^1)^{\epsilon_1} \ldots (dx^n)^{\epsilon_n} d\theta_{i_1}^{k_{i_1}} \ldots d \theta_{i_{n-p}}^{k_{i_{n-p}}} \delta^{(\ell_{j_1})} (d\theta_{j_1}) \ldots \delta^{(\ell_{j_p})} (d\theta_{j_p})$, where any of the $i_s$ is different from any of $j_s$, \emph{i.e.}\ a single $d\theta$ has either polynomial dependence or distributional. The homotopy we introduced above yields
\bear
(\mathpzc{h} \omega + \omega \mathpzc{h}) \alpha_p = (n - p - \deg_{dx} (\alpha_p) - \deg_{\delta (d\theta)} (\alpha_p) + \deg_{d\theta} (\alpha_p)) \alpha_p,
\eear
where now $\deg_{\delta (d\theta)} (\alpha_p) = \sum_{s=1}^{n-p} \ell_{i_{s}} $ and $\deg_{d\theta} (\alpha_p) = \sum_{s=1}^p k_{j_s}$. Restricting to cocycles one finds that $\deg_{\delta (d\theta)} (\alpha_p)  = 0$ and that $dx^{j_s} = 0$ for any $s = 1, \ldots p$ so that $\deg_{dx } (\alpha_p) = \sum_{s = 1}^{n-p} \epsilon_{i_s}$, \emph{i.e.} the remaining non-vanishing $dx$'s are those having the same indexes of the polynomnial $d\theta$'s appearing. On the other hand, one has that compatibility with the previous equation forces $\sum_{s=1}^{n-p} \epsilon_{i_s} k_{i_s} = 0$, which implies that $k_{i_s} = 0$ for any $i_s$ so that $ \deg_{d\theta} (\alpha_p) = 0$. This leads to elements of the following form $\alpha_p = f (x, \theta) dx^{i_1} \ldots dx^{i_{n-p}}\delta (d\theta_{j_1}) \ldots \delta (d\theta_{j_{p}}) \in \Omega^{n-p|p}_\mani$. It can be seen that all of these belong to the same class and, once again, they all transform as semidensities, \emph{i.e.} they transform with $\det (M) \cong \mathcal{B}er (\mani)^{\otimes 1/2}$. We defer these checks to the Appendix.  

\noindent Further, it is not hard to see that the de Rham differential $d$ is zero everywhere on the cohomology of $\omega$, since for any $p=0, \ldots n$, the de Rham differential $d : \Omega^{\bullet| p}_\mani \rightarrow \Omega^{\bullet+1|p}_\mani$ always moves out to a space where the cohomology of $\omega$ is zero. \\

\noindent We recollect these results in the following Theorem. 
\begin{theorem}[Cohomology $H_\omega (\Omega_\mani^{\bullet | p})$] \label{teo1} Let $\mani = \Pi \mathcal{T}^\ast M$ be an odd symplectic supermanifold of dimension $n|n$ and let $(\Omega_\mani^{\bullet|p}, d, \omega)$ for $p = 0, \ldots n$ be its extended de Rham double complex. Then the following are true:
\begin{enumerate}
\item $E_1 = H_\omega (\Omega^{\bullet|p}_\mani)$ is generated by $ \sum_{i_{1}, \ldots, i_n}\epsilon_{i_1 \ldots i_n} dx^{i_1} \ldots dx^{i_{n-p}} \delta (d\theta_{i_{n-p+1}}) \ldots \delta (d\theta_{i_{n}})$ over $\mathcal{O}_\mani$ 
In particular, 
$
\dim_{\stsheaf} H_\omega (\Omega^{\bullet|p}_\mani) = 1 
$ so that $\dim_{\stsheaf} \sum_{p=0}^n H_\omega (\Omega^{\bullet| p}_\mani) = n.$
\item Let $[\alpha_p] \in H_\omega (\Omega^{\bullet|p}_\mani)$ for any $p = 1, \ldots n.$ Then $ \langle [\alpha_p] \rangle_{\stsheaf}$ is isomorphic to semidensities on $\mani.$ 
\item The de Rham differential vanishes on all the cohomology of $\omega$.\ In particular $E_2 = H_d H_\omega (\Omega_\mani^{\bullet |p}) = E_1$ for any $p=0,\ldots, n$.
\end{enumerate}
\end{theorem}
\noindent As already said above, the case $p=0$ has been discussed by \v{S}evera \cite{Severa}. Also, notice that in the Theorem we have chosen $\sum_{i_{1}, \ldots, i_n}\epsilon_{i_1 \ldots i_n} dx^{i_1} \ldots dx^{i_{n-p}} \delta (d\theta_{i_{n-p+1}}) \ldots \delta (d\theta_{i_{n}})$ as representative for symmetry reason, but - as explained above - we might have chosen just one element in this sum.

\subsection{Lagrangian Submanifolds and Cohomology} In this subsection we want to outline a connection of the previous results with Lagrangian submanifolds in an odd symplectic supermanifold $\mani$. 
Notice that since the term sub-supermanifolds is quite clumsy, we refer at them as submanifolds instead. In particular, one defines a \emph{Lagrangian submanifold} $\mathpzc{L}$ of $\mani$ to be one that is \emph{maximally isotropic}. This means that, given an embedding $\iota : \mathpzc{L} \hookrightarrow \mani$, we require that $\iota^\ast \omega = 0$ and that $\mathpzc{L}$ is \emph{not} a proper submanifold of any other isotropic submanifold of $\mani$. Once again, the geometry of these special submanifolds has been characterized by Schwarz. 
\begin{theorem}[Schwarz \cite{Schwarz}] Let $ ( \mani, \omega )$ be an odd symplectic supermanifold, with reduced manifold $\manir$ and let $\mathpzc{L}$ be a Lagrangian submanifold in $\mani.$ Then there exists a pair $(\mathpzc{L}_{red}, \varphi )$, where $\mathpzc{L}_{red} \subset \manir$ is an ordinary manifold and $\varphi: \mani \stackrel{\sim}{\rightarrow} \Pi \mathcal{T}^\ast \manir$ is the  symplectomorphism of Theorem \ref{11}, such that $\varphi \lfloor_{\mathpzc{L}} : \mathpzc{L} \stackrel{\sim}{\rightarrow}  \mbox{\emph{Tot}} (\Pi \mathcal{T}^\ast_{\manir / \mathpzc{L}_{red}}),$ where $\Pi \mathcal{T}^\ast_{\manir / \mathpzc{L}_{red}} \subset \Pi \mathcal{T}^\ast_{\manir}$ is the odd conormal bundle of $\mathpzc{L}_{red}$.
\end{theorem} 
\noindent This theorem says that - once again up to global symplectomorphisms - the geometry of a Lagrangian submanifold in an odd symplectic supermanifold is constrained to be the one of the total space of a conormal bundle in $\mbox{Tot} (\Pi \mathcal{T}^\ast_{\manir}).$ Notice that given an ordinary manifold $M$ the odd conormal bundle of a submanifold $\iota : L \hookrightarrow M$ is defined to be the bundle whose fibers are forms which give zero on vectors in $\Pi \mathcal{T}_L$, \emph{i.e.}\  $\Pi \mathcal{T}^\ast_{M/L}$ fits in the (split, in the smooth case) short exact sequence
\bear
\xymatrix{
0 \ar[r] & \Pi \mathcal{T}^\ast_{M/L} \ar[r] & \iota^\ast (\Pi \mathcal{T}^\ast_M) \ar[r] & \Pi \mathcal{T}^\ast_L \ar[r] & 0.
}
\eear 
Since the rank of the odd conormal bundle corresponds to the odd dimension of the related Lagrangian submanifold $\mathpzc{L}$, this constraints any Lagrangian submanifold in $\mani$ to be of dimension $k| n-k$ if $M$ is of dimension $n$ (hence $\mani$ of dimension $n|n$) and $L$ is of dimension $k$. For example, let us consider the example of an odd symplectic supermanifold of dimension $(2|2)$ given by $\mani = \mbox{Tot} (\Pi \mathcal{T}_{\mathbb{R}^2})$ with its standard odd symplectic form $\omega = \sum_{i=1}^2 dx^i d\theta_i$, and look at its Lagrangian submanifolds. The first corresponds to the reduced manifold, $\manir \cong \mathbb{R}^2$, \emph{i.e.}\ it is given by the locus $\mathpzc{L}_1 \defeq \{ \theta_1 = \theta_2 = 0 \} \subset \mani,$ which singles out a $2|0$-dimensional (super)manifold. The second is given by the \virgolette complement'' of $\mathpzc{L}^{2|0}$ in $\mani$, that is $\mathpzc{L}^{0|2} \defeq \{ x_1 =x_2 = 0 \} \subset \mani$, which is a $0|2$-dimensional supermanifold. Then we have two obvious $1|1$-dimensional sub supermanifolds, which are given by the loci $\mathpzc{L}^{1|1}_1 \defeq \{ x_1=\theta_2 = 0\}$ and $\mathpzc{L}^{1|1}_2 \defeq \{ x_2=\theta_1 = 0\}$ in $\mani,$ notice by the way that this two Lagrangian submanifolds are connected by a symplectomorphism. In particular, for $t \in [0,1]$, the transformation $G_t = \mbox{diag} (A(t), A^\intercal(t)) \in SO(2) \times SO(2)$ with 
\bear
A(t) = \left ( \begin{array}{cc} 
\cos (\frac{\pi}{2} t) & -\sin (\frac{\pi}{2} t) \\
\sin (\frac{\pi}{2} t) & \cos (\frac{\pi}{2} t) 
\end{array}
\right )
\eear
\noindent rotates the Lagrangian $\mathpzc{L}^{1|1}_2$ to $\mathpzc{L}^{1|1}_1$. We say that two Lagrangians are \emph{Lagrangian-homotopic} if there exists a smooth family of Lagrangians $\mathpzc{L}_t$ for $t \in [0,1]$ that relate them. This is just the case above: the flow along $G_t$ for $t \in [0,1]$ defines a smooth family $\mathpzc{L}_t$ of Lagrangian submanifolds such that $\mathpzc{L}_{t=0} = \mathpzc{L}^{1|1}_2$ and $\mathpzc{L}_{t=1} = \mathpzc{L}^{1|1}_1.$  This defines an equivalence relation, we will write $\mathpzc{L}_1 \sim \mathpzc{L}_2$ and denote the corresponding class with $[\mathpzc{L}]$. The above example can be generalized to any $\mathbb{R}^n$, though it is fair to stress that, in general, classification of Lagrangian submanifolds is a delicate and difficult problem, intimately related to topology, and we will not dwell any further in it for it is not the aim of the present paper. \\
\noindent Now let $\Pi \mathcal{T}_\mathpzc{L}$ be the parity reversed tangent bundle of a certain Lagrangian submanifold. It can be seen that one has a splitting $\Pi \mathcal{T}_\mani \cong \Pi \mathcal{T}_\mathpzc{L} \oplus \Pi \mathcal{T}_{\mathpzc{L}^\vee}$, for a certain \emph{Lagrangian complement} $\Pi \mathcal{T}_{\mathpzc{L}^\vee}$ of $\Pi \mathcal{T}_\mathpzc{L}$ in $\Pi \mathcal{T}_\mani$. In this instance, the related Berezinian bundles, \emph{i.e.} $\mathcal{B}er (\mathpzc{L}) \defeq \mathcal{B}er (\Pi \mathcal{T}^\ast_{\mathpzc{L}})^\ast$ and $\mathcal{B}er (\mathpzc{L}^\vee) \defeq \mathcal{B}er (\Pi \mathcal{T}^\ast_{\mathpzc{L}^\vee})^\ast$ are isomorphic - notice by the way that this is not at all trivial, see \cite{Severa} for a discussion. On the other hand one has that $\mathcal{B}er (\mani) \cong \pi^{\ast} (\det (M))^{\otimes 2}$ and the splitting yields $\mathcal{B}er (\mani) \cong \mathcal{B}er (\mathpzc{L}) \otimes \mathcal{B}er (\mathpzc{L}^\vee)$, so it can be concluded that $\mathcal{B}er (\mathpzc{L}) \cong \pi^\ast (\det (M))$ for \emph{any} Lagrangian submanifold $\mathpzc{L}$ embedded into $\mani.$ This means that any Lagrangian submanifold in an odd symplectic supermanifold $\mani$ is such that its Berezinian transforms as the square root of the Berezinian of $\mani$, \emph{i.e.}\ as a semidensity on $\mani$.\\
Restricting to the case of contractible spaces or simply $\mathbb{R}^n$, by the above considerations one finds that the class $[\alpha_p] \in H_\omega (\Omega^{\bullet |p}_\mani)$ corresponds to the tensor density of the class of Lagrangian submanifolds $[\mathpzc{L}^{n-p|p}]$ of dimension $n-p|p$, which is indeed a semidensity; conversely to the class $[\mathpzc{L}^{n-p|p}]$ one associates the corresponding class into $H_\omega (\Omega^{\bullet |p}_\mani)$. An immediate example is given by the class of $[\mathpzc{L}^{n|0}] \leftrightarrow [dx^1 \ldots dx^n]$, corresponding to the Lagrangian submanifold $\manir \subset \mani$, and by its complement in $\mani$, which is a purely odd supermanifold (over a point) of dimension $0|n$ and whose class is $[\mathpzc{L}^{0|n}] \leftrightarrow [\delta (d\theta_1) \ldots \delta (d\theta_n)]$. Notice that one has $[\mathpzc{L}^{n|0}] \otimes [\mathpzc{L}^{0|n}] \cong [\mani]$, \emph{i.e.} the tensor product of two semidensities - which are Lagrangian complement of each other - yields a density for the odd symplectic supermanifold $\mani$, as explained above.

\section{Picture Changing Operators and Lagrangian Submanifolds}

\noindent Interesting objects in the context of the extended de Rham complex introduced in the previous sections are the the so-called \emph{picture changing operators} (PCO's for short), which have been introduced first in a supergeometric context by Belopolsky in \cite{Belo} in relation to issues arising in superstring perturbation theory. As their name suggests, and as briefly discussed in the introduction, these are operators that allow one to move \virgolette vertically'' (or diagonally) in the diagram \eqref{12} above. More in particular, as they act on sections of $\Omega^{\bullet| p}_\mani$, they either increase or decrease the picture $p$ and can only be characterized as {local} operators related to a certain direction singled out by a (usually constant) vector field in $\mathcal{T}_\mani$. They are better behaved when looked as operators related to the cohomology with respect to some differential, indeed they are constructed as to respect cohomology classes, \emph{i.e.}\ they maps cocycles to cocycles and coboundaries to coboundaries. The interested reader can find details in the recent \cite{Catenacci:2018xsv}, where PCO's with respect to the de Rham differential $d$ are discussed. In the present section we are interested in constructing and interpreting the PCO's with respect to $\omega$ seen as a differential as above. Inspection of the non-trivial cohomology classes suggests how to define a \emph{picture raising operator}. Namely working in a coordinate chart given by $x^i | \theta_i$ for $i = 1, \ldots, n$, one defines
\bear \label{PCOY}
\xymatrix@R=1.5pt{
\mathbb{Y} : \Omega_\mani^{\bullet | p} \ar[r] & \Omega_\mani^{\bullet -1 | p +1} \\
\alpha \ar@{|->}[r] & \mathbb{Y} (\alpha ) \defeq \sum_{i,j=1}^n \delta (d\theta_i) \delta_i^j \iota_j \alpha,
}
\eear 
where $\delta^j_i$ is the Kronecker symbol and it stands for a (conventional) choice of the direction of the PCO. In other words, the action of the picture raising operator $\mathbb{Y}$ is to replace $dx^i$'s with the corresponding $\delta(d\theta_i)$'s. Notice in particular that $\mathbb{Y}$ is an even derivative of $\Omega^{\bullet | p}_\mani$ for any $p$ and, remarkably, it preserves cohomology classes (indeed it can be checked that it commutes with $\omega$), so that one obtains a map $\mathbb{Y} : H_\omega (\Omega^{\bullet |p}_\mani) \rightarrow H_\omega (\Omega^{\bullet |p+1}_\mani).$ In particular acting with the \virgolette product'' of $n$ picture raising operators one gets a map $\prod_{i=1}^n \mathbb{Y} = \mathbb{Y}^n : H_\omega (\Omega^{\bullet |0}_\mani) \rightarrow H_\omega (\Omega^{\bullet | n}_\mani)$ from the cohomology of differential forms to the cohomology of integral forms, with $[ dx^1 \ldots dx^n ] \mapsto [ \delta (d\theta_1) \ldots \delta (d\theta_n) ]$. 
The definition of $\mathbb{Y}$ includes a conventional choice of a direction, that in \eqref{PCOY} is encoded in the Kronecker symbol. Let us clarify this point: instead of using the identity matrix, we could ask whether it is possible to make a different choice in the definition \eqref{PCOY}, namely
\begin{equation}\label{PCOYAB}
	\mathbb{Y} \defeq \sum_{i,j=1}^n \delta \left( d \theta_i \right) J_i^j \iota_j \ ,
\end{equation}
for a certain matrix $J$ (that a priori can be non-constant). If we want \eqref{PCOYAB} to be compatible with the cohomology of $\omega$, we have to require that it commutes with $\omega$, namely $[\mathbb{Y}, \omega ]= 0$, which implies
\begin{equation}\label{PCOYAC}
	 \sum_{i,j,k,l=1}^n \delta \left( d \theta_i \right) J_i^j \iota_j \left( dx^k \omega^l_{\; k} d \theta_l \right) = \sum_{i,j,l=1}^n \delta \left( d \theta_i \right) J_i^j \omega^l_{\; j} d \theta_l = 0 \ .
\end{equation}
If we consider $\omega^l_j = \delta^l_j$, then equation \eqref{PCOYAC} simply implies that $J$ is diagonal, and hence it has $n$ free entries, exactly as a vector $V$. Therefore, we can trade the definition \eqref{PCOYAB} for the equivalent
\begin{equation}\label{PCOYAD}
	\mathbb{Y}_{V} \defeq \sum_{i,j=1}^n \delta \left( V^i d \theta_i \right) V^j \iota_j \ ,
\end{equation}
where we have specified the dependence on $V$ in order to distinguish it from \eqref{PCOYAB}. It is clear that \eqref{PCOYAD} is a non-trivial map $\mathbb{Y}_{V} : H_\omega (\Omega^{\bullet|p}_\mani) \to H_\omega (\Omega^{\bullet|p+1}_\mani )$, since it commutes with $\omega$, really
\begin{equation}
	\mathbb{Y}_{V} \omega = \sum_{i,j,k=1}^n \delta \left( V^i d \theta_i \right) V^j \iota_j dx^k  d \theta_k = \sum_{i,j=1}^n \delta \left( V^i d \theta_i \right) V^j d \theta_j = 0 \ .
\end{equation}
It is easy to observe that different choices of the vector $V$ in \eqref{PCOYAD} or, equivalently, different choices of the (diagonal) matrix in \eqref{PCOYAB}, lead to cohomologically equivalent PCO's. We postpone this discussion to the example in the Appendix. We will use \eqref{PCOY} as the definition of $\mathbb{Y}$, unless differently specified.

\noindent Analogously, \emph{picture lowering operators} can be introduced. Working again in the chart $x^i | \theta_i$ for $i = 1, \ldots, n$, one defines, using the definition of $\Theta (\iota^i)$ given in \cite{Catenacci:2018xsv} 
\bear
\xymatrix@R=1.5pt{
\mathbb{Z} : \Omega_\mani^{\bullet | p} \ar[r] & \Omega_\mani^{\bullet +1 | p -1} \\
\alpha \ar@{|->}[r] & \mathbb{Z} (\alpha ) \defeq -i \omega \wedge \left ( \Theta ( V_i \iota^{i}) (\alpha) \right ),
}
\eear
where $V_i$ is a constant vector encoding the conventional choice of a direction. When not differently specified, we choose $V_i = (1, \ldots , 1)$. The presence of $\Theta (\iota^i)$\footnote{The factor $-i$ in the definition is related to the integral representation of $\Theta$, and should not distract.} makes the action of this operator less manifest compared to the picture raising operator defined above. Let us consider as an example its action on $\alpha_n = f (x, \theta) \delta (d\theta_1) \ldots \delta (d\theta_n)$, and for simplicity let us consider $V_i = (0, \ldots, 0,1,0, \ldots,0)$, i.e. the only nontrivial entry is the $i$-th:
\begin{align}
\mathbb{Z} \left ( f \left( x , \theta \right)  \delta (d\theta_1) \ldots \delta (d\theta_n) \right ) 
& = f \left ( x , \theta \right) \omega \wedge \left ( \frac{1}{d \theta_i} \delta (d\theta_1)\ldots \widehat{\delta (d\theta_i)}\ldots \delta (d\theta_n) \right )   \nonumber \\
& = f (x, \theta) dx^i \delta (d\theta_1)\ldots \widehat{\delta (d\theta_i)}\ldots \delta (d\theta_n), 
\end{align}
where $\hat{\delta}$ represents a delta not appearing in the expression: in other words, the action of the operator $\mathbb{Z}$ makes the $i$-th delta drop and it get substituted with the corresponding $dx^i$.  Notice also that in the previous calculation appears the formal expression $1/d\theta$, which in the language of string theory belongs to the \emph{large Hilbert space} (see again \cite{Catenacci:2018xsv}), but it drops at the ends of the calculation. Just like $\mathbb{Y}$ above, also $\mathbb{Z}$ is even and it preserves cohomology classes - once again it commutes with $\omega$. In analogy with what above, in particular, we have that $\prod_{i=1}^n \mathbb{Z} = \mathbb{Z}^n : H_\omega (\Omega^{\bullet | n}_\mani ) \rightarrow H_\omega (\Omega^{\bullet | 0}_\mani)$, maps the cohomology of integral forms to the cohomology of superforms, \emph{i.e.} $[\delta (d\theta_1) \ldots \delta (d\theta_n)] \mapsto [dx^1 \ldots dx^n]$. 
Finally, it follows easily that $\mathbb {Z} \mathbb{Y} = \mathbb{Y} \mathbb{Z} = 1$ in cohomology. \\

\noindent The above discussion, together with the considerations carried out in the previous sections about Lagrangian submanifolds in odd symplectic supermanifolds, yields a very nice geometrical interpretation of the action of the PCO's. We have indeed seen that forms $[\alpha]$ in the cohomology $H_\omega (\Omega_\mani^{\bullet |p})$ are related to inequivalent Lagrangian submanifolds in $\mani$: \emph{the action of the PCO's makes one jumps from one Lagrangian submanifold to another (inequivalent) Lagrangian submanifold.} In other words, PCO's does not act, geometrically, as symplectomorphisms of a Lagrangian submanifold in $\mani$: 
from this point of view, they can be seen as non-trivial maps from the cohomology $H_\omega (\Omega^{\bullet | \bullet}_\mani)$ to itself, or analogously as maps between classes of Lagrangian submanifolds of different codimension in $\mani$. \\

\noindent 
The PCO's $\mathbb{Z}$ and $\mathbb{Y}$ defined above are related to the operator (multiplication by) $\omega$. We now aim at establishing a relation between these and the PCO's related to de Rham differential $d$ and constructed as to act on the cohomology classes $H_d (\Omega^{\bullet|p}_\mani)$, see \cite{Catenacci:2018xsv}. In particular, in order to carry out this comparison, we will consider the following PCO's for $d$: the picture raising operator $\mathbb{Y}^d$ is a multiplicative operator, formally written as $\mathbb{Y}^d \defeq \sum_{i=1}^n \theta_i \delta (d\theta_i)$, whilst the picture lowering operator is given by $\mathbb{Z}^{d} \defeq -i [d, \Theta (V_i \iota^i)]$, see again \cite{Catenacci:2018xsv} for a discussion. Notice in particular that $\mathbb{Y}^d$ and $\mathbb{Z}^d$ raises and lowers the picture of the form by one respectively, whilst they leave the form number unchanged.

\noindent In the following of this section we will write the PCO's related to $\omega$ introduced above as $\mathbb{Y}^\omega$ and $\mathbb{Z}^\omega$, in order to distinguish them from those related to $d$.\\
Before we start, following the relation between the operator $d$ and the total operator $d + \omega$ 
observed in \cite{Khuda}, one can note that 
\begin{eqnarray}
\label{newPCOB}
D \defeq d + \omega = e^{\Sigma} d e^{-\Sigma}\,, \quad\quad
I_i \defeq \iota_i - \theta_i = e^{\Sigma} \iota_i e^{-\Sigma}\,, 
\end{eqnarray}
where $\Sigma = \sum_{i=1}^n dx^i \theta_i$ is the \emph{Liouville form}, which is defined so that  $\omega = - d \Sigma = \sum_{i=1}^n dx^i  d\theta_i$. 
In particular, the fact that the odd symplectic form $\omega$ is $d$-exact implies that $D^2 =0$. In addition, it is easy to check that 
$\{ I_i, I_j\} =0$ and 
\begin{eqnarray}
\label{newPCOD}
D I_i + I_i D = {\mathcal L}_i -  ({\mathcal L}_i \Sigma)\,, 
\end{eqnarray}
where $\mathcal{L}_i $ is the Lie derivative along the coordinate vector field in the related direction. Notice that this can also be derived using the similarity transformation. 
By using $\left[ d , \Sigma \right] = - \omega$ and $\left[ \omega , \Sigma \right] = 0$, one can compute (we omit the sum symbols) 
\begin{eqnarray}
\label{newPCOC}
\mathbb{Z}^d  - \mathbb{Z}^\omega = -i \left[d - \omega, \Theta(V_i \iota^i)\right] = -i
\left[ e^{-\Sigma} d e^{\Sigma}, \Theta(V_i \iota^i) \right] = -i
e^{-\Sigma} \left[d , \Theta(V_i \iota^i) \right] e^{\Sigma} = e^{-\Sigma} {\mathbb{Z}}^{d} e^{\Sigma} \ ,
\end{eqnarray}
since $[\Sigma,\Theta(V_i \iota^i)]=0$. Analogously
we have 
\begin{eqnarray}
\label{newPCOF}
\mathbb{Y}^\omega - \mathbb{Y}^d = \left[\iota_i - \theta_i, \delta(d\theta_i)\right] = 
\left[ e^{\Sigma} \iota_i e^{-\Sigma}, \delta(d\theta_i) \right] = 
e^{\Sigma} \left[ \iota_i , \delta(d\theta_i) \right] e^{-\Sigma} = e^{\Sigma} {\mathbb Y}^{\omega} e^{-\Sigma} \,. 
\end{eqnarray}
Now, similarly as above, defining ${\mathcal R} = \sum_{i=1}^n \left( \partial^i \iota_i + \partial_i \iota^i \right)$, one can verify that $[{\mathcal R}, \omega ] = d$ and $\left[ \mathcal{R} , d \right] = 2 \Delta_{BV}$. This leads in turn to $d + \omega + 2 \Delta_{BV} = e^{\mathcal R} \omega e^{-\mathcal R}$ and $\theta_i - \iota_i = e^{\mathcal R} \theta_i e^{-\mathcal R}$ so that, using $\mathcal{R}$ instead of $\Sigma$ one sees that  
\begin{align}\label{newPCOG}
\mathbb{Z}^d  - \mathbb{Z}^\omega =  e^{\mathcal{-R}} \mathbb{Z}^\omega e^{ \mathcal{R}}, \quad \quad \mathbb{Y}^\omega - \mathbb{Y}^d = e^{\mathcal{R}} \mathbb{Y}^d e^{-\mathcal{R}},
\end{align}
where we have implicitly used $\left[ \mathcal{R} , \Theta \left( V_i \iota^i \right) \right] = 0$ and $\left[ \Delta_{BV} , \Theta \left( V_i \iota^i \right) \right] = 0$. Together with \eqref{newPCOC} and \eqref{newPCOF}, one has $e^{-\Sigma} {\mathbb{Z}}^{d} e^{\Sigma} = e^{\mathcal{-R}} {\mathbb{Z}}^{\omega} e^{\mathcal{R}}$ and $e^{\Sigma} {\mathbb Y}^{\omega} e^{-\Sigma} = e^{\mathcal{R}} \mathbb{Y}^{d} e^{-\mathcal{R}}$, so that 
\begin{align}\label{newPCOH}
\mathbb{Z}^d = (e^{\Sigma} e^{\mathcal{-R}}) \mathbb{Z}^\omega (e^{\mathcal{R}}e^{-\Sigma}) = e^{ \Sigma - \mathcal{R}} \mathbb{Z}^\omega e^{-\Sigma + \mathcal{R}}, \quad \quad \mathbb{Y}^d = (e^{-\mathcal{R}} e^{\Sigma}) \mathbb{Y}^\omega (e^{- \Sigma} e^{\mathcal{R}}) = e^{\Sigma - \mathcal{R}} \mathbb{Y}^\omega e^{- \Sigma + \mathcal{R}},
\end{align}
where we note that the product of the exponential can be re-written as a unique exponential using the Baker-Campbell-Hausdorff formula, and the last equality follows from the commutation relation between $\Sigma$ and $\mathcal{R}$. Notice that the previous can interpreted as similarity transformations between the two pairs of PCO's. As a consistency check, we notice that \eqref{newPCOH} is compatible with $\mathbb{Y} \mathbb{Z} = \mathbb{Z} \mathbb{Y} = 1$ (in cohomology), which is valid for the PCO's relative to $d$ and for those relative to $\omega$.

\section{BV Laplacian on Semidensities}

\noindent In the previous sections we have explained that to Lagrangian submanifolds are attached semidensities. In this context, for example, \v{S}evera treated the special case of the reduced manifold $\manir$, which is a Lagrangian submanifold in $\mani$. Further, \v{Severa} showed also that the invariance of the BV Laplacian introduced by Khudaverdian in \cite{Khuda1} and \cite{Khuda2} follows from first principles. This operator is a certain differential - actually the second differential, in our convention and the third in \v{S}evera's - of the spectral sequence related to the double complex $(\Omega^{\bullet |0}_\mani, \omega, d)$. Something is to be stressed before we go on, though, for the situation is peculiar. We take $\omega$ to be the vertical differential and $\delta_1 \defeq d$ to be the horizontal differential of the double complex, and defining $(E_{r}, \delta_r )$ the spectral sequence related to the vertical filtration, \emph{i.e.}\ we first compute the cohomology with respect to $\omega$, so that $E_{1} = H_\omega (\Omega^{\bullet |0}_\mani)$: we have seen that when acting on $E_1$, the de Rham differential $\delta_1 = d$ is the zero map. This might lead to think that the spectral sequence converges already at page one, but this is not the case. Indeed, one finds that the differential $\delta_2$ acting on $E_2 = H_d H_\omega (\Omega_\mani^{\bullet |0}) = H_\omega (\Omega^{\bullet |0}_\mani) = E_1$ is not zero, and it is the discovery of \v{S}evera in \cite{Severa} that $\delta_2$ is indeed the BV Laplacian, $\Delta_{BV} = \sum_i \partial_{x^i} \partial_{\theta_i}$, which he writes formally as $\delta_2 = d \circ \omega^{-1} \circ d$ to get the right \virgolette movement'' for the second differential of the spectral sequence, $\delta_2 : E_{2}^{\bullet, \bullet} \rightarrow E_{2}^{\bullet +2,\bullet -1}$. This means that one finds naturally an invariant differential acting on semi-densities (generated by the element $[dx^1 \ldots dx^n] \in E_1$), corresponding to the first page $E_1$ of the spectral sequence, which is nothing but the BV Laplacian.\\
The previous analysis can be generalized to our extended framework considering $\Omega^{\bullet|p}_{\mani}$ for any $p$, and the action of the second differential $\delta_2$ on the related cohomology groups. We start reviewing the $p=0$ case.
One has that since $d$ maps elements in $H_\omega (\Omega^{\bullet |0}_\mani) $ to $\omega$-exact elements, \emph{i.e.}
\bear \label{exac}
d \left (f (x, \theta) dx^1 \ldots dx^n \right ) = \omega \alpha_0^{(n-1)},
\eear
where $\Omega^{n-1|0}_\mani \owns \alpha_0^{(n-1)} = (-1)^{|f|+1} \sum_i (\partial^i f) \iota_i (dx^1 \ldots dx^n)$. This is convenient, since the $\omega$ appearing in the \eqref{exac} and the $\omega^{-1}$ coming from $\delta_2$ cancel one another, leaving with $d\alpha_0^{(n-1)}$ to compute: 
\begin{align}
\delta_2 (f (x, \theta) dx^1 \ldots dx^n) &= (-1)^{|f| +1} \sum_{i, j = 1}^n dx^j (\partial_{j} \partial^i f (x, \theta)) \iota_i (dx^1 \ldots dx^n) \nonumber \\
& = \sum_{i=1 }^n \partial_i \partial^i f (x, \theta) (dx^1 \ldots dx^n)  + \left (\omega\mbox{-exact terms} \right ).
\end{align}
It follows that restricting the action of $\delta_2 = d \circ \omega^{-1} \circ d $ on the cohomology of $\omega$, \emph{i.e.}\ on semidensities attached to the Lagrangian submanifold corresponding to the reduced manifold, one finds indeed 
\bear
\delta_2 \left ( f(x, \theta ) [dx^1 \ldots dx^n] \right ) = \sum_{i=1}^n \frac{\partial^2}{\partial x^i \partial \theta_i} f (x, \theta) [dx^1 \ldots dx^n].
\eear
Now we show that this extends easily to any class in the full cohomology $H_\omega (\Omega_\mani^{\bullet |p})$ for any $p$. Let us start from integral forms: once again, the first observation is that 
\bear
d (f (x, \theta) \delta (d\theta_1) \ldots \delta (d\theta_n)) = \omega \alpha_n^{(-1)}
\eear
for $\Omega^{-1|n}_\mani \owns \alpha^{(-1)}_n = - \sum_{i} (\partial_i f ) \iota^i (\delta (d\theta_1) \ldots \delta (d\theta_n)$, and where we have used that $- d\theta_i \delta^{\prime} (d\theta_i) = \delta (d\theta_i).$ So, just like above, we are left to compute $d\alpha^{(-1)}_n$ from $\delta_2 (f \delta (d\theta_1) \ldots \delta(d\theta_n)) $, and one gets  
\begin{align}
\delta_2 (f (x, \theta) \delta (d\theta_1) \ldots \delta(d\theta_n)) & = - \sum_{i, j =1}^n d\theta_j \partial^j \partial_i f (x, \theta) \iota^i (\delta (d\theta_1) \ldots \delta (d\theta_n)) \nonumber \\
& = \sum_{i=1}^n ( \partial^i \partial_i f ) (\delta (d\theta_1) \ldots \delta (d\theta_n)) + \left (\omega\mbox{-exact terms} \right ).
\end{align}
Once again, this means that
\bear
\delta_2 \left ( f(x, \theta ) [\delta (d\theta_1) \ldots \delta (d\theta_n) ] \right ) = \sum_{i=1}^n \frac{\partial^2}{\partial x^i \partial \theta_i} f (x, \theta) [\delta (d\theta_1) \ldots \delta (d\theta_n)].
\eear
Finally, let us consider the case of pseudo forms: as usual these represent the case where the most attention is required, as it is a mixture of the previous situations for superforms and integral forms. In particular, adopting the notation of the previous sections, considering the representative of a class $[	\alpha_p] \in H_{\omega} (\Omega^{\bullet |p}_\mani )$ at picture $p$ given by $f_{a_1 \ldots a_n}(x, \theta) dx^{a_1} \ldots dx^{a_{n-p}} \delta (d\theta_{a_{n-p+1}}) \ldots \delta (d\theta_{a_n})$ one first observe that
\begin{align}
d \alpha_p = \omega \sum_{i=1}^{n-p} (-1)^{|f|+1}\left  ( \partial^{a_i} f_{a_1 \ldots a_n} (x, \theta) \right ) \iota_{a_i} \left ( dx^{a_1} \ldots dx^{a_{n-p}} \delta (d\theta_{a_{n-p+1}}) \ldots \delta (d\theta_{a_n}) \right )  + \nonumber \\
- \omega \sum_{i = n-p+1}^{n} \left ( \partial_{a_i} f_{a_1 \ldots a_n} (x, \theta) \right )  \iota^{a_i} \left ( dx^{a_1} \ldots dx^{a_{n-p}} \delta (d\theta_{a_{n-p+1}}) \ldots \delta (d\theta_{a_n}) \right ). 
\end{align}
Therefore, cancelling the $\omega$ with the $\omega^{-1}$ coming from $\delta_2$, one is left with 
\begin{align}
\delta_2 & \left ( f_{a_1 \ldots a_n}(x, \theta) dx^{a_1} \ldots dx^{a_{n-p}} \delta (d\theta_{a_{n-p+1}}) \ldots \delta (d\theta_{a_n}) \right )  = \\
& = (-1)^{|f|+1} \sum_{i=1}^{n-p} dx^{a_i} \left ( \partial_{a_i} \partial^{a_i} f_{a_1 \ldots a_n} (x, \theta) \right ) \iota_{a_i} \left ( dx^{a_1} \ldots dx^{a_{n-p}} \delta (d\theta_{a_{n-p+1}}) \ldots \delta (d\theta_{a_n}) \right ) \nonumber \\
& - \sum_{i=n-p+1}^{n} d\theta_{a_i} \partial^{a_i} \partial_{a_i} f_{a_1\ldots a_n} (x, \theta) \iota^{a_i} \left (dx^{a_1} \ldots dx^{a_{n-p}} \delta (d\theta_{a_{n-p+1}}) \ldots \delta (d\theta_{a_n}) \right ) + \ldots
\end{align}
where the ellipses stand for $\omega$-exact terms. Summing up the pieces one has, up to exact terms
\begin{align}
\delta_2 & \left ( f_{a_1 \ldots a_n}(x, \theta) dx^{a_1} \ldots dx^{a_{n-p}} \delta (d\theta_{a_{n-p+1}}) \ldots \delta (d\theta_{a_n}) \right )  = \nonumber \\
& \left (  \sum_{i = 1}^{n}  \partial_i \partial^i f_{a_1 \ldots a_n} (x, \theta) \right )  dx^{a_1} \ldots dx^{a_{n-p}} \delta (d\theta_{a_{n-p+1}}) \ldots \delta (d\theta_{a_n})
\end{align}
which shows that also in this case one finds 
\begin{align}
\delta_2 & \left ( f_{a_1 \ldots a_n}(x, \theta ) [dx^{a_1} \ldots dx^{a_{n-p}} \delta (d\theta_{n-p+1}) \ldots \delta (d\theta_n) ] \right ) = \nonumber \\
& \sum_{i=1}^n \frac{\partial^2}{\partial x^i \partial \theta_i} f_{a_1 \ldots a_n} (x, \theta) [dx^{a_1} \ldots dx^{a_{n-p}} \delta (d\theta_{n-p+1}) \ldots \delta (d\theta_n) ].
\end{align}
Now, one should evaluate the cohomology of the differential $\delta_2 = \Delta_{BV}$. The usual approach employed in literature is to map the BV Laplacian to the de Rham differential by means of the so-called \emph{odd Fourier transform} \cite{Mnev}, thus concluding - under suitable hypotheses - that the cohomology is just given by $\mathbb{R}$; this approach is understood in \cite{Severa}. We will follow another approach, namely we directly compute the cohomology of $\Delta_{BV}$ by showing its homotopy operator, without making use of the odd Fourier transform: among other things, this provides one with the form of the representatives of this cohomology.\\
In particular, we claim that given a (local) section of the structure sheaf of $\mani$, which we write $s = f^I(x) \theta_I$ for some multi-index $I$, then  
\bear
f^{I} \theta_I \longmapsto h (f^I \theta_I) \defeq \sum_{a=1}^n \left (\int_{0}^1 dt \,t^{Q_s} x^a G^{\ast}_t f^I (x)\right ) \otimes \theta_a \theta_I,
\eear
where $G^{\ast}_t f (x) = f(tx)$ for $t \in [0,1]$, \emph{i.e.}\ it is the pull-back of the section $s$ under the map $x \stackrel{G_t}{\longmapsto} t x$ and $Q_s$ is a constant depending on the section, to be determined later on. The tensor product is there for notational convenience (actually, one might have written $s = f^I (x) \otimes \theta_I$ from the very beginning). One has that 
\bear
h \circ \Delta_{BV} (f^I (x) \theta_I) = \sum_{a,b=1}^{n} \left ( \int_0^1 dt \, t^{Q_{\delta s}} x_b G^{\ast}_t (\partial_{x^a} f^I (x)) \right ) \otimes \theta_b \partial_{\theta_a} \theta_I.
\eear
On the other hand, one computes
\begin{align}
\Delta_{BV} \circ h (f^I (x) \theta_I) & = \sum_{a=1}^n \int_0^1 dt \, t^{Q_s} G_{t} f^I(x) \otimes \theta_I + \\
& - \sum_{a=1}^n \int_0^1dt \, t^{Q_s} G^\ast_t f^I (x) \otimes \theta_a \partial_{\theta_a} \theta_I + \\
& + \sum_{a=1}^n \int_0^1 dt \, t^{Q_s} x^a \partial_{x^a} (G^\ast_t f^I(x)) \otimes \theta_I + \\
&  - \sum_{a,b=1} \int_0^1 dt \, t^{Q_s} x_b \partial_{x^a} (G^\ast_t f^I (x)) \otimes \theta_b \partial_{\theta_a } \theta_I
\end{align}
Let us examine the various summands. Clearly the first one is just $n \int_0^1 dt \, t^{Q_s} G_{t} f^I(x) \otimes \theta_I$, while the last term cancels with the term $h \circ \Delta_{BV}$ above, upon using the chain-rule and posing $Q_{\delta s} = Q_s + 1.$ The second one can be rewritten as
\bear
\sum_{a=1}^n \int_0^1dt \, t^{Q_s} G^\ast_t f^I (x) \otimes \theta_a \partial_{\theta_a} \theta_I = \deg (\theta_I) \, \int_0^1dt \, t^{Q_s} G^\ast_t f^I (x) \otimes  \theta_I,
\eear
where $\deg (\theta_I)$ counts the number of the theta's in the monomial and it spans from $0$ to $n$. The third term can be rewritten as follows
\begin{align}
\sum_{a=1}^m \int_0^1 dt \, t^{Q_s} x^a \partial_{x^a} (G^\ast_t f^I(x)) \otimes \theta_I  = \int^{1}_0 dt \, t^{Q_s +1} \frac{d}{dt} f^I(tx) \otimes \theta_I.
\end{align}
Integrating by parts, one gets
\begin{align}
\sum_{a=1}^m \int_0^1 dt \, t^{Q_s} x^a \partial_{x^a} (G^\ast_t f^I(x)) \otimes \theta_I = f^I (x) \theta_I - \delta_{Q_s + 1,0} f(0)^I \theta_I - (Q_s + 1) \int_0^1 dt \, t^{Q_s} G^{\ast}_t f^I (x) \otimes \theta_I.
\end{align}
Altogether, one has 
\begin{align}
\left ( \Delta h + h \Delta_{BV}) (f^I (x)\theta_I \right ) & =  f^I (x) \theta_I - \delta_{Q_s + 1,0} f(0)^I \theta_I \nonumber \\
& + (n - \deg (\theta_I) - Q_s -1 ) \int_0^1 dt \, t^{Q_s} G^{\ast}_t f^I (x) \otimes \theta_I.
\end{align}
In order to cancel the last term one must set $Q_s = n - \deg (\theta_I) - 1$. In this case, one gets
\bear
\left ( \Delta_{BV} h + h \Delta_{BV}) (f^I (x)\theta_I \right ) & =  f^I (x) \theta_I - \delta_{n - \deg (\theta_I),0} \, f(0)^I \theta_I,
\eear 
therefore one gets a homotopy whenever $\deg (\theta_I) < n$. In the case $\deg (\theta_I) = n$ it is easy to see that the only cocycles are given by elements of the form $c \cdot \theta_1 \ldots \theta_n$, for $c \in \mathbb{R}$. Inserting the generating sections $[\alpha_p] \in E_2 = H_{\omega} (\Omega^{\bullet |p}_\mani),$ one sees that $E_3 = H_{\Delta_{BV}} (\Omega^{\bullet |p}_\mani) = \mathbb{R} \cdot [\theta_1 \ldots \theta_n \cdot \alpha_p]$. Notice also that, since $\theta_1 \ldots \theta_n$ transforms as $\det (M)^{\otimes -1}$, one concludes that for any $p$, the representative $\theta_1 \ldots \theta_n \cdot \alpha^{\mathpzc{L}}$ is actually invariant. Finally, notice that $c \cdot \theta_1 \ldots \theta_n \cdot \alpha_{p}$ for $c\in \mathbb{R}$ is $d$-closed. It follows that any higher differential $\delta_{i >2}$ is zero and the spectral sequence converges  to $E_3 = H_{\Delta_{BV}} (\Omega^{\bullet |p}_\mani)$, which is then isomorphic to $n$ copies of $\mathbb{R}.$ \\

\noindent 
The above discussion can be related to the other spectral sequence which arises from the double complex having $\omega$ and $d$ as differential, namely the one starting with $d$ instead of $\omega$. In order to distinguish the two spectral sequence we denote $E^d$ the one having $d$ as vertical differential and $E^\omega$ the one having $\omega$ as vertical differential. By recalling that $\omega$ is $d$-exact, it is not hard to see that $E^d$ converges already at page 1 so that $E^d_1 = H_d (\Omega^{\bullet|p}_\mani)$. In particular, one finds that $H_d (\Omega^{\bullet|p}_\mani) \cong \mathbb{R}$ for any $p = 1, \ldots, n$, so that the two spectral sequences converge indeed to the same space, isomorphic to $n$ copies of $\mathbb{R}$ whereas all the pictures are taken into account. With reference to the cohomology of $d$, the non-trivial classes are generated (over $\mathbb{R}$) by the elements $ \sum_{i_j} \theta_{i_1} \ldots \theta_{i_p} \delta (d\theta_{i_1}) \ldots \delta (d\theta_{i_p})$ for $i_j= 1, \ldots, n$, $p=0, \ldots, n$ and $i_j \neq i_k$, where the case $p=0$ corresponds indeed to the representative $1$. Notice that here, just like in the cohomology of $\omega$ above, we might have chosen a single element instead of the sum above as a representative of $H_d (\Omega^{\bullet|p}_\mani)$ for all $0<p<n$, since all of the elements in the sum are actually cohomologous - this fact can be proven in exactly the same way as we have done in the Appendix for the cohomology of $\omega$ -, nonetheless this more \virgolette democratic'' choice looks the most suitable to us. Notice that all the classes in $E_1^d = H_d (\Omega^{\bullet|p}_\mani)$ can be obtained from $1$ upon using the picture raising operators related to $d$, \emph{i.e.}\ $\mathbb{Y}^d$, whose definition is recalled in section 4. Moreover, something which is really worth stressing is that the de Rham cohomology above contains an element which is also present in $E_3 = H_{\Delta_{BV}} (\Omega^{\bullet | p}_\mani)$ discussed above, namely the only element coming from the complex of integral forms at picture $p=n$, \emph{i.e.} $\theta_1 \ldots \theta_n \delta (d\theta_1) \ldots \delta (d\theta_n):$ this will prove useful in what follows. 

\noindent Finally, we address the relation between the (isomorphic) convergence spaces of the two spectral sequences, namely $E^\omega_3= H_{{\Delta_{BV}}} (\Omega^{\bullet|p}_\mani)$ and $E^d_1 = H_d (\Omega^{\bullet|p}_\mani)$. It is indeed possible to find an explicit map, carrying the representatives of one space to the other. For a generic element $f (x, \theta , dx, d\theta) \in \Omega^{\bullet | p}_\mani$ let us define the following integral transformation
\bear \label{fodd}
(\mathcal{F}^{odd} f) (x, \theta , dx , d\theta) \defeq \int [d\eta^1 \ldots d\eta^n | d \xi_1 \ldots d \xi_n] e^{\left( \sum_{i=1}^n  \eta^i \theta_i + \sum_{j=1}^n \xi_j dx^j \right)} f (x, \xi, \eta , d\theta)
\eear
where the symbol $[d\eta_1 \ldots d\eta_n | d \xi_1 \ldots d \xi_n]$ indicates that we are Berezin-integrating along the odd coordinates $\eta$'s and the $\xi$'s, \emph{i.e.}\ along \emph{all} the odd \virgolette coordinates''. The $\eta$'s and the $\xi$'s are paired with their natural duals, the $\theta$'s and the $dx$'s: here it is worth to remember that we have identified $\theta_i = \partial_{dx^i}$, hopefully clarifying this duality. Also, notice that in the case of odd symplectic supermanifolds the symbol $[d\eta_1 \ldots d\eta_n | d\xi_1 \ldots d\xi_n]$ is invariant, so that the above is well-defined.\\
Let us now consider the action of the integral transform $\mathcal{F}^{odd}$ on a generic representative of the cohomology $E^d_1 = H_d (\Omega^{\bullet | p}_\mani)$. We set $\omega (x, \theta, dx, d\theta) = \theta_{i_1} \ldots \theta_{i_p} \delta (d\theta_{i_1}) \ldots \delta (d\theta_{i_p})$ and we compute 
\begin{align}
( \mathcal{F}^{odd} \omega ) (x, \theta, dx, d\theta) = \int  [d^n\eta | d^n \xi ] e^{\left( \sum_{i=1}^n  \eta^i \theta_i + \sum_{j=1}^n \xi_j dx^j \right)}  \xi_{i_1} \ldots \xi_{i_p} \delta (d\theta_{i_1}) \ldots \delta (d\theta_{i_p}).
\end{align}
The integral can be factorized into two Berezin integrals, 
\begin{align}
( \mathcal{F}^{odd} \omega ) (x, \theta, dx, d\theta) & = \left ( \int  [d^n\eta ] e^{\sum_{i=1}^n {\eta^i \theta_i}} \right ) \left ( \int [d^n\xi ] e^{\sum_{j=1}^n \xi_j dx^j }  \xi_{i_1} \ldots \xi_{i_p} \right )\delta (d\theta_{i_1}) \ldots \delta (d\theta_{i_p}) \nonumber \\
& = \theta_1 \ldots \theta_n \left ( \int [d^n\xi ] e^{\sum_{j=1}^n \xi_j dx^j }  \xi_{i_1} \ldots \xi_{i_p} \right )\delta (d\theta_{i_1}) \ldots \delta (d\theta_{i_p})
\end{align}
The second integral yields 
\begin{align}
\int [d^n\xi ] e^{\sum_{j=1}^n \xi_j dx^j }  \xi_{i_1} \ldots \xi_{i_p} = dx^{i_{p+1}} \ldots dx^{i_n}.
\end{align}
Putting all of the pieces back together, one gets that
\begin{align}
\xymatrix{
H_d (\Omega^{\bullet |p}_\mani) \owns \theta_{i_1} \ldots \theta_{i_p} \delta (d\theta_{i_1}) \ldots \delta (d\theta_{i_p})  \ar@{|->}[r]^{\mathcal{F}^{odd}\qquad \quad}& \theta_1 \ldots \theta_n dx^{i_{p+1}}\ldots dx^{i_{n}}  \delta (d\theta_{i_1}) \ldots \delta (d\theta_{i_p}) \in H_{\Delta_{BV}} (\Omega^{\bullet | p}_\mani)
}.
\end{align}
The inverse map is easily figured out as the anti-transform of the previous, and it maps $H_{\Delta_{BV}} (\Omega^{\bullet | p}_\mani)$ to $H_d (\Omega^{\bullet |p}_\mani)$. \\

\noindent We show that it is easier to relate these cohomologies by using both the sets of PCO's for $d$ and for $\omega$ defined previously. As observed above, a hint comes from the fact that the element 
\begin{equation}
	\beta = c \cdot \theta_1 \ldots \theta_n \cdot \delta \left( d \theta_1 \right) \ldots \delta \left( d \theta_n \right), \quad  c \in \mathbb{R} \ .
\end{equation}
belongs to both of the cohomologies $H_{d} (\Omega^{\bullet |p}_\mani)$ and $H_{\Delta_{BV}} (\Omega^{\bullet | p}_\mani)$: one can then use this elements as \virgolette pivot'' and acts on it with the picture changing operators of $d$ and $\omega$ as to get \emph{any} elements of both the cohomologies. In particular, any representative in $E^\omega_3 = H_{\Delta_{BV}} (\Omega^{\bullet |p}_\mani)$ can be obtained by applying certain powers of the picture lowering operator $\mathbb{Z}^{\omega}$ to $\beta$:
\begin{equation}\label{BVAAAA}
	c \cdot \theta_1 \ldots \theta_n \cdot \alpha^{(n-k|k)} = \left[ \prod_{i=1}^{n-k} \mathbb{Z}^{\omega} \right] \beta \ \Longleftrightarrow \ \beta = \left[ \prod_{i=1}^{n-k} \mathbb{Y}^{\omega} \right] c \cdot \theta_1 \ldots \theta_n \cdot \alpha^{(n-k|k)} \ .
\end{equation}
Analogously, any representative found in $E^d_1 = H_{\Delta_{BV}} (\Omega^{\bullet |p}_\mani)$ can be obtained by applying certain powers of the picture lowering operator $\mathbb{Z}^{d}$ to $\beta$:
\begin{equation}\label{BVAAAB}
	c \cdot \alpha^{(0|k)} = \left[ \prod_{i=1}^{n-k} \mathbb{Z}^{d} \right] \beta \ \Longleftrightarrow \ \beta = \left[ \prod_{i=1}^{n-k} \mathbb{Y}^{d} \right] c \cdot \alpha^{(0|k)} \ .
\end{equation}
By confronting \eqref{BVAAAA} and \eqref{BVAAAB} we get the identity
\begin{equation}
	\left[ \prod_{i=1}^{n-k} \mathbb{Y}^{\omega} \right] c \cdot \theta_1 \ldots \theta_n \cdot \alpha^{(n-k|k)} = \left[ \prod_{i=1}^{n-k} \mathbb{Y}^{d} \right] c \cdot \alpha^{(0|k)} \ ,
\end{equation}
or analogously
\begin{equation}
	c \cdot \theta_1 \ldots \theta_n \cdot \alpha^{(n-k|k)} = \left[ \prod_{i=1}^{n-k} \mathbb{Z}^{\omega} \mathbb{Y}^{d} \right] c \cdot \alpha^{(0|k)}, \qquad  c \cdot \alpha^{(0|k)} = \left[ \prod_{i=1}^{n-k} \mathbb{Z}^{d} \mathbb{Y}^{\omega} \right] c \cdot \theta_1 \ldots \theta_n \cdot \alpha^{(n-k|k)} \ .
\end{equation}
We can evaluate explicitly these expression, thus obtaining the simple equations
\begin{equation}
	c \cdot \theta_1 \ldots \theta_n \cdot \alpha^{(n-k|k)} = \left[ \prod_{i=1}^{n-k} \theta_{a_i} dx^{a_i} \right] c \cdot \alpha^{(0|k)}, \qquad  c \cdot \alpha^{(0|k)} = \left[ \prod_{i=1}^{n-k} \partial^{a_i} \iota_{a_i} \right] c \cdot \theta_1 \ldots \theta_n \cdot \alpha^{(n-k|k)} \ .
\end{equation}
This leads to the following interpretation: we can use the PCO's as \virgolette ladder operators" in order to move from representatives of the cohomology of $d$ to representatives of the cohomology of $\Delta_{BV}$ and viceversa, once again establishing an isomorphism between the two cohomologies, as the following diagram explains pictorially 
\bear
\xymatrix{
\ar@/^.5pc/[ddr]^{\mathbb{Y}^d} H_{d} (\Omega^{\bullet|p}_\mani) \ar@/^.5pc/[rr] &  &\ar@/^.5pc/[ll] H_{\Delta_{BV}} (\Omega^{\bullet |p}_\mani) \ar@/^.5pc/[ddl]^{\mathbb{Y}^\omega} \\ \\
& \ar@/^.5pc/[uul]^{\mathbb{Z}^d} [\theta_{1} \ldots \theta_n \delta(d\theta_1)\ldots \delta(d\theta_n) ]. \ar@/^.5pc/[uur]^{\mathbb{Z}^\omega} & 
}
\eear
We recollect the result of this section in the following Theorem, which is in some sense the completion of Theorem \ref{teo1} above.
\begin{theorem}[BV Cohomology and de Rham Cohomology] Let $\mani $ be an odd symplectic supermanifold of dimension $n|n$ and let $(\Omega_\mani^{\bullet|p}, d, \omega)$ for $p = 0, \ldots n$ be its extended de Rham double complex, let $E_\bullet^\omega$ and $E_\bullet^d$ be the spectral sequences starting with $\omega$ and $d$ respectively. Then the following are true.
\begin{enumerate}
\item $E_{\bullet}^\omega \Rightarrow E_3^{\omega} = H_{\Delta_{BV}} (\Omega^{\bullet |p}_\mani) $ for $p=0, \ldots, n$. This cohomology is generated over the real numbers by $ [\theta_1 \ldots \theta_n \cdot \alpha_p]$ for $\alpha_{p}$ a generator of $H_\omega (\Omega^{\bullet | p}_\mani)$, as in Theorem \ref{teo1}, and it is isomorphic to $\mathbb{R} $ for any $p$. \\
In particular, the homotopy of the BV operator $\Delta_{BV} \defeq \sum_{i=1}^n \partial_{x^i}\partial_{\theta_i} $ acting on local sections $f^I \theta_I$ of $\mathcal{O}_\mani$ for some multi-index $I$ is given by 
\bear
h (f^I \theta_I) \defeq \sum_{a=1}^n \left (\int_{0}^1 dt \,t^{Q_s} x^a G^{\ast}_t f^I (x)\right ) \otimes \theta_a \theta_I,
\eear
where $G^{\ast}_t f (x) = f(tx)$ for $t \in [0,1]$ and $Q_s = n-1 - \deg(\theta_I)$.
\item $E_{\bullet}^d \Rightarrow E_1^{d} = H_d (\Omega^{\bullet |p}_\mani)$ for $p=0,\ldots, n.$ This cohomology is generated over the real numbers by $[\sum_{i_j} \theta_{i_1} \ldots \theta_{i_p} \delta (d\theta_{i_1}) \ldots \delta (d\theta_{i_p})]$ for $i_j \neq i_k$, $p = 0, \ldots,n$ and it is isomorphic to $\mathbb{R}$ for any $p$.
\end{enumerate}
Finally, an explicit isomorphism between the cohomologies $H_{\Delta_{BV}} (\Omega^{\bullet | p}_\mani)$ and $H_{d} (\Omega^{\bullet |p}_\mani)$ is realized by using the PCO's for $\omega$ and for $d$ or via the map $\mathcal{F}^{odd}$ in \eqref{fodd} and its inverse. 
\end{theorem}

\section{New PCO's from Old and Analogies with Kodaira-Spencer-Type Theory}

\subsection{New Picture Changing Operators from Old}

\noindent In this section we introduce new picture changing operators that are built starting from those described in the section 3 above. In particular, we define new \emph{odd} picture changing operators, in contrast with those defined above. We will show that these new operators can be used to define the BV operator, but also different BV-type operators that changes the picture number as well.  
In particular, we define the operator $\displaystyle \check{d} $ via the picture changing operator $\mathbb{Y}^\omega$, that we have introduced above:
\bear \label{NPCOA}
\xymatrix@R=1.5pt{
	 \check{d} : \Omega_\mani^{\bullet | p} \ar[r] &  \Omega_\mani^{\bullet | p+1}  \\
	 \alpha \ar@{|->}[r] & \check{d} \alpha \defeq \left[ d , \mathbb{Y}^{\omega} \right] \alpha,
}
\eear
where we note that the operator $\check{d}$ raises the picture number by one, but it leaves the form degree invariant: in other words, it moves vertically in the \virgolette stack'' of complexes in \eqref{12} above. Also, notice that this operator is nilpotent. In analogy, we define the operator $\displaystyle \hat{d} $ using $\mathbb{Z}^\omega$
\bear \label{NPCOB}
\xymatrix@R=1.5pt{
	 \hat{d} : \Omega_\mani^{\bullet | p} \ar[r] &\Omega_\mani^{\bullet +2| p-1}  \\
	 \alpha \ar@{|->}[r] & \hat{d} \alpha \defeq \left[ d , \mathbb{Z}^{\omega} \right] \alpha.
}
\eear
Notice that $\hat d$ does not move vertically, but in oblique direction: it lowers the picture by one and raises the form degree by two.

\noindent Let us now recover the BV Laplacian introduced above, by means of $\hat d$ and $\check d$. Let us consider the operator 
\begin{equation}\label{NPCOC}
	\tilde{\Delta} \defeq - \frac{1}{2} \left( \check{d} \circ \left( \omega \right)^{-1} \circ \hat{d} + \hat{d} \circ \left( \omega \right)^{-1} \circ \check{d} \right) \ ,
\end{equation}
where $\omega $ is the odd symplectic form. We use the definitions of the operators in \eqref{NPCOA} and \eqref{NPCOB} to obtain, dropping the composition symbols
\begin{align}
	\nonumber \tilde{\Delta} &= - \frac{1}{2} \Big[ \left( d \mathbb{Y}^{\omega} - \mathbb{Y}^{\omega} d \right) \left( \omega \right)^{-1} \left( d \mathbb{Z}^{\omega} - \mathbb{Z}^{\omega} d \right) +  \\
	\label{NPCOD} & \quad + \left( d \mathbb{Z}^{\omega} - \mathbb{Z}^{\omega} d \right) \left( \omega \right)^{-1} \left( d \mathbb{Y}^{\omega} - \mathbb{Y}^{\omega} d \right) \Big] \ .
\end{align}
When acting on the cohomology of $\omega$, \emph{i.e.} taking an element $\alpha \in H_\omega (\Omega^{\bullet | p }_\mani ) $, it is easy to show that only two terms are not trivial and the action is exactly that of the BV Laplacian:
\begin{eqnarray}
	\label{NPCOE} \tilde{\Delta} \alpha = \frac{1}{2} \left[ \mathbb{Y}^{\omega} \Delta_{BV} \mathbb{Z}^{\omega} + \mathbb{Z}^{\omega} \Delta_{BV} \mathbb{Y}^{\omega} \right] \alpha = \Delta_{BV} \alpha .
\end{eqnarray}
The interested reader can find in the Appendix an explicit example, showing the action of \eqref{NPCOC}.


\noindent Likewise, one can define \virgolette BV-type operators" which modify the picture and the form number:
\begin{align}
	\Delta_{BV}^{(-2|2)} &\defeq \check{d} \circ \omega^{-1} \circ \check{d}, \\
	\Delta_{BV}^{(-1|1)} &\defeq \frac{1}{2} \left[ \check{d} \circ \omega^{-1} \circ d + d \circ \omega^{-1} \circ \check{d} \right], \\
	\Delta_{BV}^{(1|-1)} &\defeq \frac{1}{2} \left[ \hat{d} \circ \omega^{-1} \circ d + d \circ \omega^{-1} \circ \hat{d} \right], \\
	\Delta_{BV}^{(2|-2)} &\defeq \hat{d} \circ \omega^{-1} \circ \hat{d}.
\end{align}
Again, the action of these operators can be obtained from $\Delta_{BV}$ and the two PCO's $\displaystyle \mathbb{Y}^{\omega} $ and $\mathbb{Z}^{\omega} $, as above.

\subsection{Analogies with Kodaira-Spencer-Type Theory and Deformations}

\noindent In this section we describe an analogy with complex geometry which makes use of the picture changing operators introduced in the previous sections.


We start recalling basic facts about \emph{almost complex structures}. Given a real manifold ${M}^{2n}$ of dimension $2n$, an almost complex structure is an endomorphism of tangent space $\mathcal{J} \in End (\mathcal{T}{M^{2n}})$ such that $\mathcal{J}^2 =  -\text{id}_{\mathcal{T}{M^{2n}}}$. Locally one can represent $\mathcal{J}$ as a sum $\mathcal{J} = J + \bar {J}$, with 
\begin{equation}\label{KSB}
	\ J \defeq J^b_{\bar{a}} d \bar{z}^{\bar{a}} \otimes \partial_b, \qquad  \bar{J} \defeq \bar{J}^{\bar{b}}_a dz^a \otimes \bar{\partial}_{\bar{b}} 
\end{equation}
such that $J^a_{\bar{a}} \bar{J}^{\bar{a}}_b = - \delta^a_b$ and $J^a_{\bar{a}} \bar{J}^{\bar{b}}_a = - \delta_{\bar{a}}^{\bar{b}}$.
In particular, these can be used to transform holomorphic vectors into anti-holomorphic vectors and viceversa:
\begin{equation}\label{KSC}
	J \left( \bar{\partial}_{\bar{a}} \otimes 1 \right) = \left (J^b_{\bar{c}} d \bar{z}^{\bar{c}} ( \bar{\partial}_{\bar{a}} ) \right ) \partial_b = J^b_{\bar{a}} \partial_b \ , \ \bar{J} \left ( \partial_a \otimes 1 \right) = \big ( \bar{J}^{\bar{b}}_c dz^c \left( \partial_a \right) \big ) \bar{\partial}_{\bar{b}} = \bar{J}^{\bar{b}}_a \bar{\partial}_{\bar{b}} \ .
\end{equation}
Analogously, one can use $J$ and $\bar J$ to take holomorphic $(1,0)$-forms into anti-holomorphic $(0,1)$-forms and viceversa, upon using that $(\mathcal{T}^\ast M^{2n})^\ast \cong \mathcal{T} M^{2n}:$
\begin{equation}\label{KSD}
	J \left( 1 \otimes d z^a \right) = J^b_{\bar{a}} d \bar{z}^{\bar{a}} \otimes \partial_b \left( dz^a \right) = J^a_{\bar{a}} d \bar{z}^{\bar{a}} \ , \ \bar{J} \left( 1 \otimes d \bar{z}^{\bar{a}} \right) = \bar{J}^{\bar{b}}_a dz^a \otimes \bar{\partial}_{\bar{b}} \left( d \bar{z}^{\bar{a}} \right) = \bar{J}^{\bar{a}}_a dz^a \ .
\end{equation}
In the previous formulae, one can rewrite $J$ and $\bar J$ as 
\begin{equation}\label{KSE}
	J = J^b_{\bar{a}} d \bar{z}^{\bar{a}} \iota_b, \qquad  \bar{J} = \bar{J}^{\bar{b}}_a dz^a \bar{\iota}_{\bar{b}} \ ,
\end{equation}
where $\iota_a$ and $\bar{\iota}_{\bar{a}}$ are the contractions along a holomorphic and a anti-holomorphic base vector respectively.

\noindent By using the almost complex structure we can define the \emph{Nijehuis tensor} $N$:
\begin{equation}\label{KSF}
	N \left( U , V \right) = \left[ J (U) , J (V) \right] - J \left[ J (U) , V \right] - J \left[ U , J (V) \right] - \left[ U , V \right] 
\end{equation}
where $U, V \in \mathcal{T}M^{2n}$. The vanishing of $N$ is said \emph{integrability condition}: an almost complex structure with vanishing $N$ defines a complex structure so that the pair $(M^{2n}, \mathcal{J})$ lifts to a \emph{complex} manifold $X$. 
Remarkably, integrability of $N$ is equivalent to nilpotency of the anti-Dolbeault operator:
\begin{equation}\label{KSG}
	N = 0 \ \Longleftrightarrow \ \bar{\partial}^2 = 0.
\end{equation}
Taking this point of view, for a certain complex manifold $X$, given a $(0,1)$-form valued in $\mathcal{T}_X$, call it $A \in \Omega^{(0,1)}_X \otimes \mathcal{T}_X$, the \emph{deformations} of the complex structure of $X$ are therefore defined by the equation $(\bar \partial + A)^2 = 0$, that is
\begin{equation}\label{KSH}
	\bar{\partial} A + A \wedge A = 0. 
\end{equation}
Writing $A= d \bar{z}^{\bar{i}} A_{\bar{i}}^j \partial_j$, this reads
\begin{equation}\label{KSHA}
	\bar{\partial}A^i + A^j \partial_j A^i = 0,
\end{equation}
which is the \emph{Tian-Todorov equation} for the deformations of complex structures, see \cite{Mirror}.

\noindent We now try to make a connection with the construction of the previous section: we argue the following analogy
\begin{align}
\mbox{holomorphic forms} \quad & \longleftrightarrow \quad dx^i, \\
\mbox{anti-holomorphic forms} \quad & \longleftrightarrow \quad \delta (d\theta_i).
\end{align}
In other words, we identify the form degree with the holomorphic form degree and the picture number with the anti-holomorphic form degree in a theory over the complex numbers - notice that the $\delta (d\theta)$'s transform (at first order) exactly as a form on the underlying manifold. First, let us see how the analogue of a complex structure looks like. Using the above analogy, from \eqref{KSE}, we see that $J$ and $\bar {J}$ become a sort of non-diagonal versions of the previously introduced picture changing operators. In particular, along with this analogy, one can introduce a new picture raising operator, namely
\begin{equation}\label{KSI}
	\check{\mathbb{Y}} \defeq J^i_j \delta \left( d \theta_j \right) \iota_i,
\end{equation}
which can also be rewritten as 
$
\check{\mathbb{Y}} = J^i_j \delta \left( d \theta_j \right) \otimes \partial_i,
$
upon using the natural isomorphism $\mathcal{T}\mani \cong (\mathcal{T}^\ast \mani)^\ast.$ In particular, using this representation, we can define the action of \eqref{KSI} on vector fields:
\begin{align}
	\label{KSIA} \check{\mathbb{Y}} \left( \partial^i \otimes 1 \right) = J^j_k \left[ \delta \left( d \theta_k \right)  \left( \partial^i \right) \right] \partial_j & = J^j_k \left[ \delta \left( \iota^i \right) \delta \left( d \theta_k \right) \right] \partial_j = J^j_k \delta^k_i \partial_j = J^j_i \partial_j \ , \\
	\label{KSIAB} &\check{\mathbb{Y}} \left( \partial_i \otimes 1 \right) = 0 \ .
\end{align}
where $\delta (\iota^i)$ is defined by its Fourier representation, see for example \cite{Witten}. Similarly, the picture lowering operator reads
\begin{equation}\label{KSK}
\check{\mathbb{Z}} = J^i_j dx^j \delta \left( \iota^i \right),
\end{equation}
Given these definitions, we can construct the analogous of the Nijenhuis tensor \eqref{KSF} above, but taking $J \defeq \check{\mathbb{Z}} + \check{\mathbb{Y}}$ and where the brackets are now graded.
In particular, let us consider the action of this tensor, we call it $\check{N}$, on two \emph{odd} coordinate fields - notice that it follows from \eqref{KSIAB} that on even coordinate fields it yields zero. It can be seen that
\begin{equation}\label{KSM}
	\check N \left( \partial^i , \partial^j \right) = \left[ \mathbb{Y} (\partial^i) , \mathbb{Y} (\partial^j) \right] = \left( J_i^k \partial_k J^l_j - J_j ^k \partial_k J^l_i \right) \partial_l \ .
\end{equation}
Let us consider the case $\check{N} \left( \partial^i , \partial^j \right) = 0 $, which can be seen as the analogue condition $N (U,V) = 0$ when applied to two anti-holomorphic coordinate base vectors. It is remarkable to observe that in analogy with \eqref{KSG} we find that
	\begin{align}\label{KSN}
		\check{d}^2 = 0 \ \Longleftrightarrow \ \delta \left( d \theta_i \right) \delta \left( d \theta_j \right) J_i^k \partial_k J^l_j \partial_l = 0 \  \Longleftrightarrow \ \left( J_i^k \partial_k J^l_j - J_j^k \partial_k J^\rho_i \right) \partial_l = 0 \ ,
	\end{align}
	where the antisymmetrisation is due to $\delta \left( d \theta \right)$ being odd. In other words, the requirement of nilpotency of the operator $\check{d}$ is analogous to the requirement of vanishing of $\check{N}$ along odd directions (along even directions is automatic).
\noindent We can push forward our analogy to make contact with deformations of complex structure. In this context, we consider in particular \emph{Kodaira-Spencer theory} \cite{BCOV} which accounts for the deformations of the complex structure of a certain Calabi-Yau threefold $CY_3$. 

\noindent First of all, it is worth to observe that one can deform the equation above $\check{d}^2 = 0$ by setting $\check{d} \mapsto \check{d} + \delta \left( d \theta_i \right) A_i^j \partial_j = \check{d} + \delta A^j \partial_j$ as to get the equation
\begin{equation}\label{KSO}
	\check{d} A^i + A^j \partial_j A^i = 0 \ ,
\end{equation}
which has the same form as the equation \eqref{KSHA}. Moreover, it is possible to obtain this equation from a field theory action, in analogy with the Kodaira-Spencer action, that we briefly recall. Given a Calabi-Yau threefold $CY_3$, the action reads
\begin{equation}\label{KSP}
	S = \int_{CY_3} A' \wedge \frac{1}{\partial} \bar{\partial} A'+ A' \wedge \left( A \wedge A \right)' \ ,
\end{equation}
where $\displaystyle A\in \Omega^{(0,1)}_{CY_3} \otimes T^{(1,0)}_{CY_3}$, see \cite{BCOV} for details. The prime appearing above amounts to the map $A \mapsto A \wedge \Omega$, where $\Omega$ is the (global) holomorphic $3$-form of $CY_3,$ that is
\begin{equation}\label{KSR}
	A^\prime = \left( d \bar{z}^{\bar{i}} A_{\bar{i}}^j \partial_j \right) \left( \Omega_{r k l} d z^r \wedge d z^k \wedge d z^l \right) = A_{\bar{i}}^j \Omega_{j k l} d \bar{z}^{\bar{i}} \wedge d z^k \wedge d z^l \in \Omega_{CY_3}^{(2,1)}.
\end{equation}
As already said, the variation of the action leads to the Tian-Todorov equation \eqref{KSHA}. 

\noindent Let us exploit the analogy described above to deduce the action leading to \eqref{KSO}. In particular, instead of the field $A$, we consider the picture changing operator $\check{\mathbb{Y}} = \delta \left( d \theta_i \right) J_i^j \iota_j$. 
The map via the holomorphic 3-form described above now amounts to apply the operator $\check{\mathbb{Y}}$ to the volume form $\epsilon_{i j k}dx^i \wedge dx^j \wedge dx^k$ of the reduced manifold, hence
\begin{equation}\label{KST}
	\left( \check{\mathbb{Y}} \right)^\prime = \delta \left( d \theta_i \right) J_i^j \epsilon_{j k l} dx^k \wedge dx^l \ .
\end{equation}
In analogy with the Kodaira-Spencer action, one therefore writes 
\begin{equation}\label{KSU}
	S = \int_{\mani^{3|3}} \left( \check{\mathbb{Y}} \right)' \wedge \frac{1}{d} \check{d} \left( \check{\mathbb{Y}} \right)' + \left( \check{\mathbb{Y}} \right)' \wedge \left( \check{\mathbb{Y}} \wedge \check{\mathbb{Y}} \right)',
\end{equation}
where $\mani$ is a supermanifold of dimension $3|3.$
The equations of motion read
\begin{equation}\label{KSV}
	\check{d} \left( \check{\mathbb{Y}} \right)'+ d \left( \check{\mathbb{Y}} \wedge \check{\mathbb{Y}} \right)' = 0,
\end{equation}
which, in turn, performing the calculations, yield 
\begin{equation}
	\partial_{[i} J^j_{k]} + \partial_l J^j_{[i} J_{k]}^l = 0.
\end{equation}
These are the analogous of the Tian-Todorov equations. 

\appendix

\section{Cohomology and Transformation Properties for $n=2$}

\noindent In this appendix we aim at clarifying the constructions of the previous section in the easy but non-trivial case $n=2$. \\
Let us start from the cohomology of $\omega$: while the cases of differential and integral forms are trivial, the case of pseudo forms deserves some more attention. First let us show that the representatives $dx^1 \delta (d\theta_2)$ and $-dx^2 \delta (d\theta_1)$ in $H_\omega (\Omega^{\bullet|1}_\mani)$ are cohomologous. Let $t \in [0,1]$, consider the following family 
\bear
\phi_t  \defeq \left ( \left( 1 - t \right) dx^1 - t dx^2 \right ) \delta \left( t d \theta_1 + \left( 1-t \right) d \theta_2 \right).
\eear 
Notice that $\phi_0 = dx^{1} \delta (d\theta_2)$ and $\phi_1 = -dx^2 \delta (d\theta_1)$. A short calculation yields that for any $t \in [0,1]$ one has that $\omega \phi_t =0,$ \emph{i.e.} $\phi_t$ is closed for any $t$. Also, upon using the properties of the delta's, one finds  
\begin{align}
\label{AEEF} 
	\phi_t &= 
	dx^1 \delta \left( d \theta_2 + \frac{t}{1-t} d \theta_1 \right) - \frac{t}{1-t} dx^2 \delta \left( d \theta_2 + \frac{t}{1-t} d \theta_1 \right) \\ 
	&= dx^1 \delta \left( d \theta_2 \right) + \sum_{n=1}^{\infty} dx^1 \frac{1}{n!} \left( \frac{t}{1-t} \right)^n d \theta_1^n \delta^{(n)} \left( d \theta_2 \right) - \sum_{n=0}^\infty dx^2 \frac{1}{n!} \left( \frac{t}{1-t} \right)^{n+1} d \theta_1^n \delta^{(n)} \left( d \theta_2 \right). \nonumber 
\end{align}
The second and third terms gather in an exact term as to give
\bear
	\phi_t = dx^1 \delta \left( d \theta_2 \right) + \omega \left( \sum_{n=1}^{\infty} \frac{1}{n!} \left( \frac{t}{1-t} \right)^n d \theta_1^{n-1} \delta^{(n)} \left( d \theta_2 \right) \right ). 
\eear 
Hence, $\phi_t \equiv dx^1 \delta (d\theta_2)\, \mbox{mod} (\omega).$ Likewise, one sees that 
\bear
	\phi_t = -dx^2 \delta \left( d \theta_1 \right) - \omega \left( \sum_{n=1}^{\infty} \frac{1}{n!} \left( \frac{1-t}{t} \right)^n d \theta_2^{n-1} \delta^{(n)} \left( d \theta_1 \right) \right ), 
\eear
that is $\phi_t \equiv -dx^2 \delta (d\theta_1)\, \mbox{mod} (\omega),$ so that in turn $\phi_0 = dx^1 \delta (d\theta_2) \sim \phi_t \sim - dx^2 \delta (d\theta_1) = \phi_1,$ thus showing that the two elements are indeed cohomologous.\\ 
On the same line, this result could have been obtained also by observing that if we write a generic cohomology representative of $H_\omega ( \Omega^{\bullet|1}_\mani)$  by choosing a certain vector $V$, singling out a certain direction, $\alpha^{1|1}_V = V^i \epsilon_{ij} dx^j  \delta( V^i d\theta_i),$ then it is not hard to show that the infinitesimal variation of $\alpha^{1|1}_V$ with respect to $V$ reads $\delta \alpha^{1|1}_V = \omega \left  ( V^i \epsilon_{i j} \delta V^j \delta^{(1)}( V^i d\theta_i) \right )$. This says that the dependence on the choice of the vector $V$ is $\omega$-exact and the cohomology representative is unique.\\

Let us now address the transformation properties of the representatives in the cohomology at picture $p=1$ and show that they indeed transforms as semidensities over $\mani.$ 
Let us consider a certain change of coordinates of $\mani$. By the transformation properties of an odd symplectic supermanifold discussed in the first section of this paper, the action on forms goes as follows 
\begin{equation}\label{TPA}
	\left( \begin{matrix}
		dx^1 \\ dx^2 \\ d \theta_1 \\ d \theta_2
	\end{matrix} \right) \longmapsto \left( \begin{matrix}
		a & b & 0 & 0 \\
		c & d & 0 & 0 \\
		0 & 0 & \frac{d}{\Delta} & -\frac{c}{\Delta} \\
		0 & 0 & -\frac{b}{\Delta} & \frac{a}{\Delta}
	\end{matrix} \right) \left( \begin{matrix}
		dx^1 \\ dx^2 \\ d \theta_1 \\ d \theta_2
	\end{matrix} \right) \ ,
\end{equation}
where $\Delta = ad-bc$ is the determinant of the matrix $A$, which is nothing but the Jacobian of the change of coordinates on the reduced manifold $\manir.$ Let us now consider the representative $dx^1 \delta (d\theta_2)$, one finds (for $a \neq 0$)
\begin{align}\label{TPD}
	dx^1 \delta \left( d \theta_2 \right) \longmapsto &\left( a dx^1 + b dx^2 \right) \delta \left( \frac{a}{\Delta} d \theta_2 - \frac{b}{\Delta} d \theta_1 \right) =  \nonumber \\ 
	&= \Delta \left[ dx^1 \delta \left( d \theta_2 \right) + \sum_{n=1}^\infty \frac{\left( d \theta_1 \right)^n }{n!} \left( - \frac{b}{a} \right)^n \delta^{(n)} \left( d \theta_2 \right) + \sum_{n=0}^\infty \frac{\left( d \theta_1 \right)^n }{n!} \left( \frac{b}{a} \right)^{n+1} \delta^{(n)} \left( d \theta_2 \right) \right] \ ,
\end{align}
where we have expanded the $\delta$'s with respect to $d \theta_2$ and left the closed, non-exact term $dx^1 \delta \left( d \theta_2 \right)$ explicit. All the terms, except for the first, group together to form $\omega$-exact terms: in particular, one finds that
\begin{equation}\label{TPF}
	dx^1 \delta \left( d \theta_2 \right) \mapsto \Delta dx^1 \delta \left( d \theta_2 \right) + \omega \left ( \Delta \sum_{n=1}^\infty \frac{1}{n!} \left( - \frac{b}{a} \right)^n \left( d \theta_1 \right)^{n-1} \delta^{(n)} \left( d \theta_2 \right) \right ),
\end{equation}
which shows that $[dx^1 \delta (d\theta_2)] \mapsto \det (A) [dx^1 \delta (d\theta_2)]$ in cohomology, which concludes the proof.

\noindent Finally, we want to show the explicit calculations for the operator defined in \eqref{NPCOC}. For the sake of clarity, let us consider the action of \eqref{NPCOC} on $\alpha = f(x,\theta) dx^1  dx^2 $ (we assume for simplicity that $|f| = 0$). We have
\begin{equation}\label{AEE2A}
	\tilde{\Delta} \alpha = - \frac{1}{2} \hat{d} \circ \omega^{-1} \circ \check{d} \alpha \ ,
\end{equation}
since $\hat{d} \alpha = 0$. Let us first calculate $\check{d}\alpha$:
\begin{align}
	\nonumber \check{d} \alpha &= d \left[ - f dx^2 \delta \left( d \theta_1 \right) + f dx^1 \delta \left( d \theta_2 \right) \right] - \mathbb{Y}^{\omega} \left[ \partial^1 f d \theta_1 dx^1 dx^2 + \partial^2 f d \theta_2 dx^1 dx^2 \right] = \\
	\nonumber &= \omega \left[ \partial_1 f dx^2 \delta' \left( d \theta_1 \right) - \partial_2 f dx^1 \delta' \left( d \theta_2 \right) \right] .
\end{align}
We are therefore left with
\begin{align}
	(\hat{d} \circ \omega^{-1} \circ \check{d}) \alpha = \left( d \mathbb{Z}^{\omega} - \mathbb{Z}^{\omega} d \right) \left[ \partial_1 f dx^2 \delta' \left( d \theta_1 \right) - \partial_2 f dx^1 \delta' \left( d \theta_2 \right) \right] \ .
\end{align}
Let us show that the first term is trivially $0$. Indeed, by recalling the algebraic rules $\displaystyle -i \Theta \delta \left( d \theta \right) = {1}/{d \theta}$ and $\displaystyle -i \Theta \delta' \left( d \theta \right) = - {1}/{(d \theta)^2}$, one has
\begin{align}
 \mathbb{Z}^{\omega} \check{d} \alpha & = -i \left( \omega \Theta + \Theta \omega \right) \left[ \partial_1 f dx^2 \delta' \left( d \theta_1 \right) - \partial_2 f dx^1 \delta' \left( d \theta_2 \right) \right] = \\
	\label{AEE2B} &= \partial_1 f \frac{dx^1 dx^2}{d \theta_1} + \partial_2 f  \frac{dx^1 dx^2}{d \theta^2} - \partial_1 f  \frac{dx^1 dx^2}{d \theta_1} - \partial_2 f  \frac{dx^1 dx^2}{d \theta_2} = 0 \ .
\end{align}
We are therefore left with
\begin{equation}
	\label{AEE2C} \mathbb{Z}^{\omega} d \left( \partial_1 f dx^2 \delta' \left( d \theta_1 \right) - \partial_2 f dx^1 \delta' \left( d \theta_2 \right) \right) = $$ $$ = \mathbb{Z}^{\omega} \left( - \partial^1 \partial_1 f dx^2 \delta \left( d \theta_1 \right) + \partial^2 \partial_2 f dx^1 \delta \left( d \theta_2 \right) \right) \ .
\end{equation}
It is useful to observe that the two terms in \eqref{AEE2C} are representatives of $ H_\omega ( \Omega^{\bullet|1}_\mani ) $, hence they are $\omega$-closed: the previous equation thus reduces to
\begin{equation}
	-i \omega \Theta \left( - \partial^1 \partial_1 f dx^2 \delta \left( d \theta_1 \right) + \partial^2 \partial_2 f dx^1 \delta \left( d \theta_2 \right) \right) = \omega \left( - \partial^1 \partial_1 f \frac{dx^2}{d \theta_1} + \partial^2 \partial_2 f \frac{dx^1}{d \theta_2} \right) = $$ $$
	= \partial^1 \partial_1 f dx^1 dx^2 + \partial^2 \partial_2 f dx^1 dx^2 = \Delta_{BV} f dx^1 dx^2 \ ,
\end{equation}
hence proving the claim.

\section*{Acknowledgement}

\noindent This work has been partially supported by Universit\`a del Piemonte Orientale research funds, by Italian Ministero dell'Universit\`a e della Ricerca (MIUR), and by Istituto Nazionale di Fisica Nucleare (INFN) through the ``FieLds And Gravity'' (FLAG) and ``Gauge theories, Strings, Supergravity'' (GSS) research projects. 
S.N. would like to express his gratitude to Julia Giorgi for quite a lot of reasons. 


\begin{thebibliography}{99}

\bibitem{Belo} A. Belopolsky, \emph{Picture Changing Operators in Supergeometry and Superstring Theory}, arXiv:9706033

\bibitem{BV} I.\ A.\ Batalin, G.\ A.\ Vilkovisky, \emph{Gauge Algebra and Quantization}, Phys. Lett. B {\bf 102}, 1 (1981) 

\bibitem{Berkovits:2004px} 
  N.~Berkovits,
  \emph{Multiloop amplitudes and vanishing theorems using the pure spinor formalism for the superstring},
  JHEP {\bf 0409}, 047 (2004)



\bibitem{BCOV} M.\ Bershadsky, S.\ Cecotti, H.\ Ooguri, C.\ Vafa, \emph{Kodaira-Spencer theory of gravity and exact results for quantum string amplitudes}, Comm.\ Math.\ Phys.\ {\bf 165}, 311--427 (1994)

\bibitem{CNproj} S.\ L.\ Cacciatori, S.\ Noja, \emph{Projective Superspaces in Practice}, J.\ Geom.\ Phys., {\bf 130} (2018) 40--62 

\bibitem{CNR} S. Cacciatori, S. Noja, R. Re, \emph{Non Projected Calabi-Yau Supermanifolds over $\mathbb{P}^2$}, Math.\ Res.\ Lett.\ {\bf 26} (4) 1027-1058 (2019)

\bibitem{Fioresi} C.\ Carmeli, L.\ Caston, R.\ Fioresi, \emph{Mathematical Foundations of Supersymmetry}, EMS (2011)

\bibitem{Catenacci:2016qzd} 
  L.\ Castellani, R.\ Catenacci and P.\ A.\ Grassi,
  \emph{Integral Representations on Supermanifolds: Super Hodge Duals, PCOs and Liouville Forms},
  Lett.\ Math.\ Phys.\  {\bf 107}, 1, 167 (2017)


\bibitem{Catenacci:2018xsv} 
  R.~Catenacci, P.~A.~Grassi, S.~Noja,
  \emph{Superstring Field Theory, Superforms and Supergeometry},
  J.\ Geom.\ Phys.\  {\bf 148}, 103559 (2020)



\bibitem{CGNinf} R. Catenacci, P.A. Grassi, S. Noja, \emph{$A_\infty$-Algebra from Supermanifolds}, Ann.\ Henri Poincar\'{e} {\bf 20} (12) 4163--4195 (2019)  


\bibitem{CremoniniGrassi}
  C.\ A.\ Cremonini, P.\ A.\ Grassi,
  \emph{Pictures from Super Chern-Simons Theory}, 
 JHEP {\bf 2003} (2020) 043
  
\bibitem{CremoniniGrassi2}
  C.\ A.\ Cremonini, P.\ A.\ Grassi,
  \emph{Super Chern-Simons Theory: BV-formalism and $A_\infty$-algebras},
  arXiv:1912.10807 

\bibitem{CremoniniGrassi3}
  C.\ A.\ Cremonini, P.\ A.\ Grassi, S. Penati,
  \emph{Supersymmetric Wilson Loops via Integral Forms},
  arXiv:2003.01729


\bibitem{Mirror} K.Hori {et alii}, \emph{Mirror Symmetry}, Clay Mathematical Monograph, AMS (2003) 


\bibitem{Khuda1} H. M.\ Khudaverdian, \emph{Semidensities on Odd Symplectic Supermanifolds}, Commun. Math. Phys., {\bf 247}, 353-390 (2004)

\bibitem{Khuda2} H. M.\ Khudaverdian, \emph{Laplacians in Odd Symplectic Geometry}, Contemp. Math. {\bf 315}, 199-212 (2002)

\bibitem{Khuda} 
  H.\ M.\ Khudaverdian, Th.\ Th.\ Voronov,
  \emph{Differential forms and odd symplectic geometry},
  Geometry, Topology and Mathematical Physics. S. P. Novikov
  seminar: 2006-2007, V. M. Buchstaber and I. M. Krichever, eds. AMS
  Translations, Ser. 2, Vol. 224, Amer. Math. Soc., Providence, RI, 2008,
  159-171


\bibitem{Manin} 
  Y.~I.~Manin,
  \emph{Gauge Field Theory And Complex Geometry}, Springer (1988)

  
\bibitem{Mnev} 
  P.\ Mnev,
  \emph{Quantum Field Theory: Batalin-Vilkovisky Formalism and its Applications}, AMS (2019)
  
  
\bibitem{NCPMR} S. Noja, S. L. Cacciatori, F. Dalla Piazza, A. Marrani, R. Re, \emph{One-Dimensional Super Calabi-Yau Manifolds and their Mirrors}, JHEP {\bf 1704} (2017) 094 


\bibitem{Noja} S. Noja, \emph{Supergeometry of $\Pi$-Projective Spaces}, J.\ Geom.\ Phys., {\bf 124} (2018) 286--299 


\bibitem{Severa} 
   P. \v{S}evera, \emph{On the Origin of the BV Operator on Odd Symplectic Supermanifolds}, Lett. Math. Phys. {\bf 78}, 55-59 (2006)

  
  \bibitem{Schwarz} 
A.\ S.\ Schwarz, \emph{Geometry of Batalin-Vilkovisky Quantization}, Comm.\ Math.\ Phys.\ {\bf 155} (1993)
249-260; 

\bibitem{Voronov} Th.\ Th.\ Voronov, \emph{Geometric Integration Theory on Supermanifolds}, Cambridge Scientific Publisher (2014)

\bibitem{Witten} E. Witten, \emph{Notes on Supermanifolds and Integration}, arXiv:1209.2199

 
  

\end{thebibliography}
\end{document}